\documentstyle[11pt,epsfig]{article} 
\def\beq{\begin{equation}}
\def\eeq{\end{equation}}
\def\beqar{\begin{eqnarray}}
\def\eeqar{\end{eqnarray}}
\def\barr#1{\begin{array}{#1}}
\def\earr{\end{array}}
\def\bfi{\begin{figure}}
\def\efi{\end{figure}}
\def\btab{\begin{table}}
\def\etab{\end{table}}
\def\bce{\begin{center}}
\def\ece{\end{center}}
\def\nn{\nonumber}

\def\text{\textstyle}

\def\al{\alpha}

\def\de{\delta}
\def\eps{\varepsilon}

\def\si{\sigma}

\def\Ga{\Gamma}
\def\De{\Delta}

\def\refeq#1{\mbox{(\ref{#1})}}
\def\ucite#1{~\cite{#1}}
\def\citere#1{\mbox{Ref.~\cite{#1}}}

\def\solid{\raise.9mm\hbox{\protect\rule{1.1cm}{.2mm}}}
\def\dash{\raise.9mm\hbox{\protect\rule{2mm}{.2mm}}\hspace*{1mm}}

\newcommand{\GeV}{\unskip\,\mathrm{GeV}}
\newcommand{\MeV}{\unskip\,\mathrm{MeV}}
\newcommand{\TeV}{\unskip\,\mathrm{TeV}}

\def\mathswitchr#1{\relax\ifmmode{\mathrm{#1}}\else$\mathrm{#1}$\fi}

\newcommand{\PW}{\mathswitchr W}
\newcommand{\PZ}{\mathswitchr Z}

\newcommand{\PH}{\mathswitchr H}

\newcommand{\Pb}{\mathswitchr b}
\newcommand{\Pc}{\mathswitchr c}
\newcommand{\Pt}{\mathswitchr t}

\newcommand{\PWpm}{\mathswitchr {W^\pm}}
\newcommand{\PWO}{\mathswitchr {W^0}}

\newcommand{\Rb}{R_\Pb}
\newcommand{\Rc}{R_\Pc}
\newcommand{\Gb}{\Ga_\Pb}
\newcommand{\Gc}{\Ga_\Pc}

\newcommand{\Gh}{\Ga_{\mathrm h}}
\newcommand{\GT}{\Ga_{\mathrm T}}
\newcommand{\Gl}{\Ga_{\mathrm l}}

\def\mathswitch#1{\relax\ifmmode#1\else$#1$\fi}

\newcommand{\MW}{\mathswitch {M_\PW}}
\newcommand{\MWpm}{\mathswitch {M_\PWpm}}
\newcommand{\MWO}{\mathswitch {M_\PWO}}

\newcommand{\MZ}{\mathswitch {M_\PZ}}
\newcommand{\MH}{\mathswitch {M_\PH}}

\newcommand{\Mb}{\mathswitch {m_\Pb}}
\newcommand{\Mt}{\mathswitch {m_\Pt}}

\newcommand{\scrs}{\scriptscriptstyle}
\newcommand{\sw}{\mathswitch {s_{\scrs\PW}}}

\newcommand{\swbar}{\mathswitch {\bar s_{\scrs\PW}}}

\newcommand{\GF}{\mathswitch {G_\mu}}

\newcommand{\chidof}{\chi^2_{\mathrm{min}}/_{\mathrm{d.o.f.}}}

\newcommand{\yb}{y_\Pb}

\newcommand{\alpz}{\alpha(\MZ^2)}
\newcommand{\alps}{\alpha_{\mathrm s}}
\newcommand{\alpsz}{\alpha_{\mathrm s}(\MZ^2)}

\newcommand{\bos}{{\mathrm{bos}}}
\newcommand{\fer}{{\mathrm{ferm}}}

\newcommand{\SC}{{\mathrm{SC}}}
\newcommand{\IB}{{\mathrm{IB}}}

\newcommand{\LEP}{{\mathrm{LEP}}}
\newcommand{\SLD}{{\mathrm{SLD}}}

\newcommand{\lsim}
{\;\raisebox{-.3em}{$\stackrel{\displaystyle <}{\sim}$}\;}
\newcommand{\gsim}
{\;\raisebox{-.3em}{$\stackrel{\displaystyle >}{\sim}$}\;}

\def\lsim{\:\raisebox{-0.5ex}{$\stackrel{\textstyle<}{\sim}$}\:}
\def\gsim{\:\raisebox{-0.5ex}{$\stackrel{\textstyle>}{\sim}$}\:}
\begin{document}
\title{ELECTROWEAK THEORY AND \\ THE 1996/97 PRECISION ELECTROWEAK DATA
  \thanks{Invited talk presented at the XXI School of Theoretical
    Physics, Ustr\'{o}n, Poland, September 1997, to appear in Acta
    Physica Polonica.} 
\thanks{Supported by the Bundesministerium f\"ur
    Bildung und Forschung, Bonn, Germany, Contract 05 7BI92P (9) and the
    EC-network contract CHRX-CT94-0579.} 
} 
\author{D.\ Schildknecht\\
    Fakult\"at f\"ur Physik, Universit\"at Bielefeld, D-33501
    Bielefeld} 

\maketitle
\begin{abstract}
  We review the empirical evidence for the validity of the Standard
  Electroweak Theory in Nature. The experimental data are interpreted in
  terms of an effective Lagrangian for Z~physics, allowing for potential
  sources of SU(2) violation and containing the predictions of the
  Standard Electroweak Theory as a special case. Particular emphasis is
  put on discriminating loop corrections due to fermion-loop
  vector-boson propagator corrections on the one hand, from corrections
  depending on the non-Abelian structure and the Higgs sector on the
  other hand. Results from recently obtained fits of the Higgs-boson
  mass are reported, yielding $\MH \lsim 430$ GeV [680 GeV] at 95\%
  C.L.\ based on the input of $\swbar^2(\LEP+\SLD)_{`97}~ = 0.23152 \pm
  0.00023$ [$\swbar^2(\LEP)_{`97}~=0.23196 \pm 0.00028$]. The LEP2 data
  provide first direct experimental evidence for non-zero non-Abelian
  couplings among the electroweak vector bosons.
\end{abstract}
  
\section{Z Physics}

The spirit in which I will look at the electroweak precision data may be
characterized by quoting Feynman who once said:
\begin{quote}
'' In any event, it is always a good idea to try to see how much 
or how little
of our theoretical knowledge actually goes into the 
ana\-lysis of those situations 
which have been experimentally checked.''
\hfill
R.P. Feynman\ucite{FEY} 
\end{quote}

\subsection{The {$\alpha(0)$}-Born Prediction}
 
The quality of the data on electroweak interactions may be particularly well appreciated by
starting with an analysis in terms of the Born approximation
of the Standard Electroweak Theory (Standard Model, SM)\ucite{GLA,WEI}.
{}From the input of 
\begin{eqnarray}
\alpha (0)^{-1} & = & 137.0359895(61), \qquad
G_\mu = 1.16639(2) \cdot 10^{-5} {\rm GeV}^{-2}, \nn\\
\MZ & = & 91.1863 \pm 0.0020 {\rm GeV}, 
\label{1}
\end{eqnarray}
one may predict the partial width of the Z for decay into leptons, $\Gl$,
the weak mixing angle, $\swbar^2$, and the $W$ mass, $\MW$. 
The 
$\alpha (0)$-Born approximation, in
distinction from the $\alpha (\MZ^2)$-Born approximation to be introduced below,
\begin{eqnarray}
\swbar^2(1-\swbar^2) & = & \frac{\pi \alpha (0)}{\sqrt 2 G_\mu \MZ^2},\nn\\
\Gl & = & \frac{G_\mu M^3_Z}{24\pi \sqrt 2} \left( 1 + (1 - 4
\swbar^2)^2\right),\nn\\
\MW^2 & = & \MZ^2 (1 - \swbar^2) ,  
\label{2}
\end{eqnarray}
then yields  
\begin{equation}
\swbar^2 = 0.2121, \qquad
\Gl = 84.85~{\rm MeV}, \qquad
\MW = 80.940~{\rm GeV}.
\label{3}
\end{equation}
A comparison with the experimental data from tab.~1, 
\begin{eqnarray}
\swbar^2(\LEP+\SLD) & = & 0.23165 \pm 0.00024,  \qquad
\Gl = 83.91 \pm 0.11\MeV, \nn\\
\MW & = & 80.356 \pm 0.125\GeV, 
\label{4}
\end{eqnarray}
shows discrepancies between the $\alpha (0)$-Born approximation and the data by
many standard deviations. 
{\doublerulesep 3pt
\btab
\bce
\footnotesize
\begin{tabular}{|@{}c@{}||c|}
\hline
leptonic sector & hadronic sector \\
\hline
\hline
$\Gl = 83.91 \pm 0.11 \MeV$ & 
      $\begin{array}{rr} 
      R = 20.778~~ \pm & 0.029 \\ 
      75~~ \pm & 27 
      \end{array}$ \\
\hline
$\begin{array}{rr}
\swbar^2|_{\LEP} = 0.23200~~ \pm &0.00027 \\ 
196~~ \pm &28
\end{array}$ &     
      $\begin{array}{rr}
      \GT = 2494.6~~ \pm & 2.7 \MeV \\
      .8~~ \pm & 2.5 \phantom{\MeV} 
      \end{array}$ \\
\hline
$\begin{array}{rr}
\swbar^2|_{\SLD} = 0.23061~~ \pm & 0.00047 \\
55~~ \pm & 41\\
\end{array}$ &
      $\begin{array}{rr}
      \si_{\mathrm h} = 41.508~~ \pm &0.056 \\
      486~~ \pm &53 \\ 
      \end{array}$ \\
\hline
$\begin{array}{rr}
\swbar^2|_{\mathrm{LEP+SLD}} = 0.23165~~ \pm &0.00024 \\
 52~~ \pm & 23 \\
\end{array}$ &     $\begin{array}{rr}
      \Gh = 1743.6~~ \pm & 2.5 \MeV \\ 
 .1~~ \pm &.3 \phantom{MeV} \\
      \end{array}$ \\
\hline
$\begin{array}{rr}
\MW = 80.356~~ \pm & 0.125 \GeV \\
 430~~ \pm & 80 \phantom{\GeV}  \\
\end{array}$ &
      $\begin{array}{rr}
      \Rb = 0.2179~~ \pm & 0.0012 \\
      74~~ \pm & 9 \\
      \end{array}$ \\
\hline
&     $\Gb = 379.9~~ \pm ~~2.2 \MeV$ \\
\hline
&
      $\begin{array}{rr}
      \Rc = 0.1715~~ \pm &0.0056 \\
      27~~ \pm & 50 \\
      \end{array}$ \\
\hline
&     $\Gc = 299.0~~ \pm ~~9.8 \MeV$ \\
\hline \hline
input parameters & correlation matrices \\
\hline \hline
$\begin{array}{c}
\begin{array}{rr}
\MZ = 91.1863~~ \pm &0.0020 \GeV \\
 67~~ \pm & 20 \phantom{\GeV}
\end{array} \\
\hline
\GF = 1.16639 (2) \cdot 10^{-5} \GeV^{-2} \\ 
\hline
\alpz^{-1} = 128.89 \pm 0.09 \\ 
\hline
\alpsz = 0.123\pm0.006 \\ \hline
\Mb = 4.7\GeV \hspace{1pt} \\ \hline
\begin{array}{rr}
\Mt = 175\phantom{.6}~~ \pm &6\phantom{.5} \GeV\\
 5.6~~ \pm & 5.5 \phantom{\GeV}
\end{array}
\end{array}$
&
$\begin{array}{c}
\rule{0mm}{23.5mm}
\begin{array}[b]{|c||c|c|c|c|}
\hline
& \si_{\mathrm h} & R & \GT \\ \hline\hline
\si_{\mathrm h} & \phantom{-}1.00 & \phantom{-}0.15 & -0.14 \\ \hline
R               & \phantom{-}0.15 & \phantom{-}1.00 & -0.01 \\ \hline
\GT             & -0.14 & -0.01 & \phantom{-}1.00 \\ \hline
\end{array}
\\
\begin{array}[b]{|c||c|c|c|}
\hline
& \Rb & \Rc \\ \hline\hline
\Rb  & \phantom{-}1.00 & -0.23 \\ \hline
\Rc  & -0.23 & \phantom{-}1.00 \\ \hline
\end{array}
\end{array}$ \\
\hline
\end{tabular}
\vspace*{2mm}
\ece
\caption[]{\it
  The 1996 precision data (and below these data the last digits of the
  1997 data), consisting of the LEP data\ucite{LEPEWWG9602}, the SLD
  value\ucite{SLD,LEPEWWG9602} for $\swbar^2$, and the world 
  average\ucite{UA2,LEPEWWG9602} for
  $\MW$.  The partial widths $\Gl$, $\Gh$, $\Gb$, and $\Gc$ are obtained
  from the observables $R = \Gh/\Gl$, $\si_{\mathrm h} =
  (12\pi\Gl\Gh)/(\MZ^2\Ga^2_{\mathrm T})$, $\Rb = \Gb/\Gh$, $\Rc =
  \Gc/\Gh$, and $\GT$ using the given correlation matrices. The data in
  the upper left-hand column will be referred to as ``leptonic sector''
  subsequently. Inclusion of the data in the upper right-hand column
  will be referred to as ``all data''.  If not stated otherwise, the SM
  predictions will be based on the input parameters given in the lower
  left-hand column of the table, where $\alpz$ is taken from
  \citere{JEG}, $\alpsz$ results from the event-shape
  analysis\ucite{BET} at LEP, and $\Mt$ represents the direct Tevatron
  measurement\ucite{CDF}. Note that the difference between the 1996 and
  the 1997 data is half a standard deviation at most.}
 \label{tab:data}
\etab
}
  
\subsection{The {$\alpha(\MZ^2)$}-Born, the Full Fermion-Loop 
and the Complete One-Loop Standard Model Predictions}

Turning to corrections to the $\alpha (0)$-Born approximation, I follow the 1988
strategy
``to isolate and to test
directly the 'new physics' of boson loops and other new phenomena by comparing
with and looking for deviations from the predictions of the
dominant-fermion-loop results''\ucite{GOU}. Accordingly, let us strictly
discriminate\ucite{kn91,BKK,DKK,DKS,DSW} 
vacuum-polarization contributions due to fermion loops in the
photon, $Z$ and $W$ propagators from all other loop corrections, the
``bosonic'' loops, which contain virtual vector bosons within the loops. I note
that this distinction between two classes of loop corrections is gauge invariant
in the SU(2)$_L\times$U(1)$_Y$ electroweak theory. Otherwise the 
theory would fix the number of
fermion families. The reason for systematically
discriminating fermion loops in the propagators from the rest is in fact
obvious. The fermion-loop effects, leading to ``running'' of coupling constants
and to mixing among the neutral vector bosons, can be precisely predicted from
the {\it empirically known couplings} of the leptons and the (light) quarks,
while other loop effects, such as vacuum polarization due to boson pairs and
vertex corrections, depend on the {\it empirically unknown\footnote{\rm 
Compare, however, the most recent results on the trilinear couplings among the
vector bosons to be discussed in Section 2.}
 couplings} among the
vector bosons and the properties of the Higgs scalar. It is in fact the
difference between the fermion-loop predictions and the full one-loop results
which sets the scale\ucite{GOU} 
for the precision required for empirical tests of the
electroweak theory beyond (trivial) fermion-loop effects. One should remind
oneself that the experimentally unknown bosonic interactions are right at the
heart of the celebrated renormalizability properties\ucite{THO} of the electroweak
non-Abelian gauge theory\ucite{WEI}. 

When considering fermion loops, let us first of all look at the contributions of
leptons and quarks to the photon propagator. Vacuum polarization due to leptons
and quarks, or rather hadrons in the latter case, leads to the well-known
increase (``running'') of the electromagnetic coupling as a function of the
scale at which it is measured. While the contribution of leptons can be
calculated in a straightforward manner, the one of quarks is more
reliably obtained from the cross section for $e^+ e^-\to$ hadrons
via a dispersion relation\ucite{JEG,JEG1}. As a consequence of the 
experimental errors in this
cross section, in particular in the region below about
$3.5\GeV$, the value of the electromagnetic fine-structure constant at the $Z$
scale, relevant for LEP1 physics, contains a non-negligible error,   
\beq 
\alpha (\MZ^2)^{-1} = 128.89 \pm 0.09.
\label{5}
\eeq
Replacing $\alpha (0)$ in  
\refeq{2} by $\alpha (\MZ^2)$ implies replacing 
$\swbar^2$
in \refeq{2} by $s^2_0$,
\beq
s^2_0 (1 - s^2_0) = \frac{\pi \alpha (\MZ^2)}{\sqrt 2 G_\mu \MZ^2}, 
\label{6}
\eeq
which may be expected to be a more appropriate parameter  
for electroweak physics at the Z-boson scale
than the mixing angle from the $\alpha(0)$-Born approximation \refeq{2}. 
As the transition from $\alpha (0)$ to $\alpha (\MZ^2)$ is an
effect purely due to the electromagnetic interactions of leptons and quarks (hadrons),
even present in the absence of weak interactions, the relations 
\refeq{2} with the
replacement $s^2_W \rightarrow s^2_0$ from \refeq{6} 
may appropriately be called the ``$\alpha
(\MZ^2)$-Born approximation''\ucite{NOV} of the electroweak theory. 

\begin{figure}
\begin{center}
\begin{picture}(15,10.5)
\put(0.1,8){$\bar\sw^2$}
\put(1.7,0.6){$\MWpm/\MZ$}
\put(9.4,0.9){$\Gl/\MeV$}
\put(-1.8,-4.5){\includegraphics{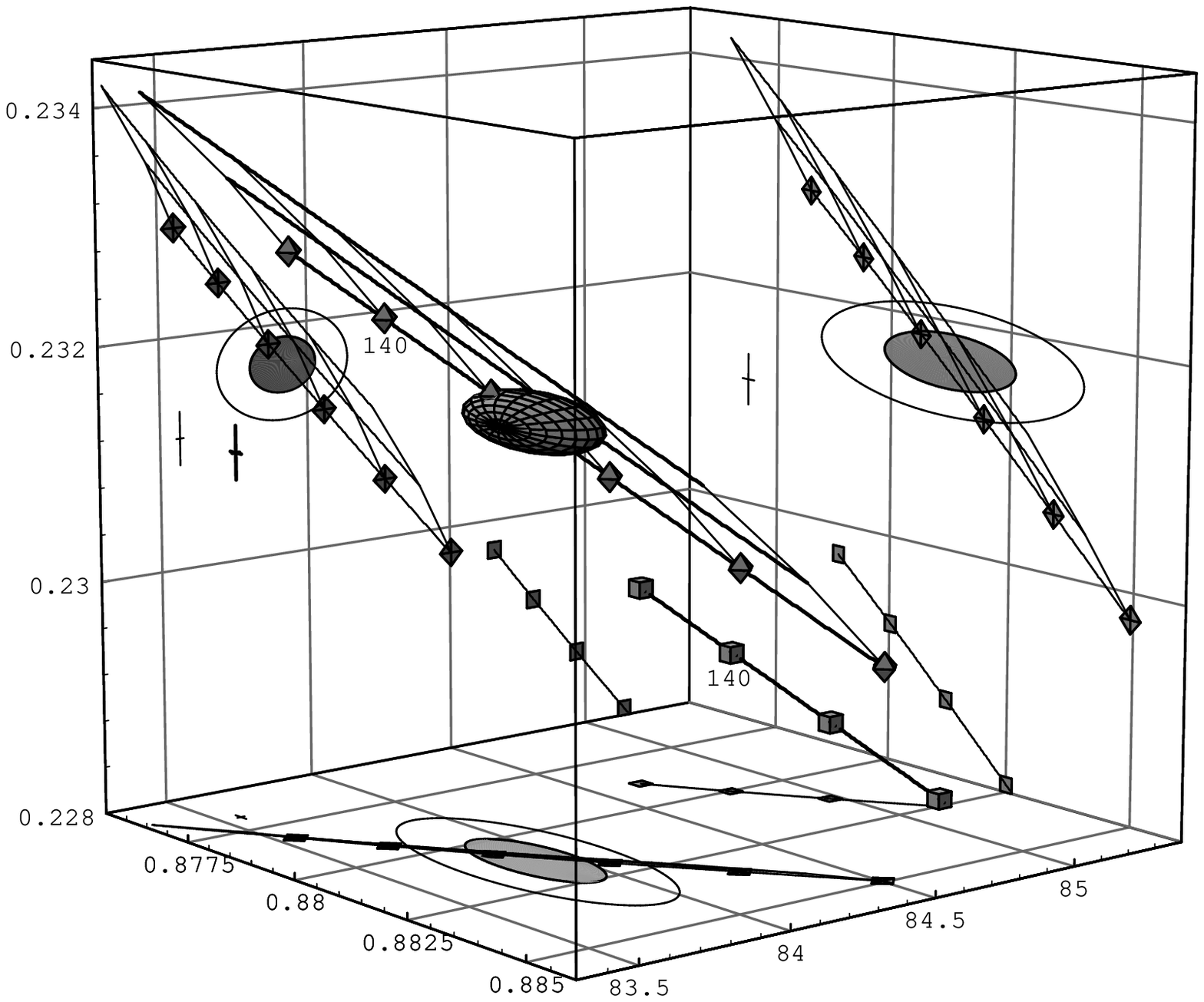}}
\end{picture}
\end{center}
\caption[]{\it Three-dimensional plot of the $1\sigma$ ellipsoid
of the 1996 experimental data in ($\MWpm/\MZ$, $\bar\sw^2$, $\Gl$)-space, using
$\bar\sw^2$ (LEP + SLD) as experimental input for $\swbar^2$, in 
comparison with the full SM prediction (connected lines)
and the pure fermion-loop prediction (single line with cubes).
The full SM prediction is shown for Higgs-boson masses of $\MH = 100
\GeV$ (line with diamonds), $300\GeV$, and 
$1\TeV$ parametrized by $\Mt$
ranging from 120--220$\GeV$ in steps of $20\GeV$. In the pure
fermion-loop prediction the cubes also indicate steps in $\Mt$ of
$20\GeV$ starting with $\Mt = 120$ $\GeV$.
The cross outside the ellipsoid indicates the $\al(\MZ^2)$-Born
approximation with the corresponding error bars,
which also apply to all other SM predictions
(1996 update from \citere{DSW}). Note that in the projections on the
planes also 
the $2\sigma$ contours are shown.  }
\label{sm3d}
\efi
Numerically, one finds 
\beqar
s^2_0 & = & 0.23112 \pm 0.00023 ,            \nn \\
\Gl^{(0)} & = & 83.563 \pm 0.012~{\rm MeV} ,    \nn      \\
M^{(0)}_{\PW} & = & 79.958 \pm 0.011~{\rm GeV} ,   
\label{6a}
\eeqar 
i.e.\ a large part of the discrepancy between 
the predictions \refeq{3} and the data 
\refeq{4} is due
to the use of the inappropriate 
value of $\alpha (0)$, instead of $\alpha (\MZ^2)$, as
appropriate for Z~physics. Note that the uncertainty in $s^2_0$, as a
consequence of the error in $\alpha (\MZ^2)$, is as large as the error of 
$\swbar^2$ from the measurements at the Z~resonance (compare \refeq{4} or
tab.~1). 

All other fermion-loop effects are due to fermion loops in the W
propagator (relevant simce $G_\mu$ enters the predictions) and in the Z propagator,
and due to the important effect of $\gamma$Z mixing induced by fermions. 
Light fermions as well as the
top quark accordingly yield important contributions to the ``full fermion-loop prediction''
which includes {\it all} fermion-loop propagator corrections. 

In fig.~1, an update of a figure in \citere{DSW}, 
we show the experimental data from the 
``leptonic sector'', $\bar
s^2_W , \Gl , \MW$, in comparison with the $\alpha (\MZ^2)$-Born
approximation, the full fermion-loop prediction, and the complete one-loop
Standard Model results. Note that fig.~1 shows the 1996 data. According to
tab.~1, the difference between the 1997 data [4]
 and the 1996 data is much below
one standard deviation and irrelevant for the content of fig.~1 and 
most of the further conclusions.\\
We conclude that\ucite{DSW,DKK},
\begin{itemize}
\item[i)]
contributions beyond the $\alpha (\MZ^2)$-Born approximation are needed for
agreement with the data, 
\item[ii)]
contributions beyond the full fermion-loop predictions, based on $\alpha(\MZ^2)$,  
the
fermion-loop contributions to the $W$ and $Z$ propagators 
and to $\gamma Z$ mixing, and the
top quark effects, are necessary, and provided
\item[iii)]
by additional contributions involving bosonic loops, dependent on the
non-Abelian couplings and the properties of the Higgs boson. 
\end{itemize}

The increase in precision of the experimental data may be particularly
well appreciated by comparing with the results which I discussed at the
XVII International School of Theoretical Physics in Szczyrk, in September
1993\ucite{Szczyrk}.

The question immediately arises of what can be said in more detail about the
various contributions due to fermionic and bosonic loops, leading to the final
agreement between theory and experiment.   
\newpage
\subsection{Effective Lagrangian, {$\Delta x, \Delta y, 
\varepsilon , \Delta\yb$} Parameters}
 
This question can be answered by an analysis in terms of the parameters 
$\Delta x, \Delta y$ and $\varepsilon$ which within the framework of an effective 
Lagrangian\ucite{BKK,DKK,DKS} 
specify potential sources of SU(2) violation. 
The {\it ``mass parameter''} $\De x$ is related to SU(2) violation by
the masses of the triplet
of charged and 
neutral (unmixed) vector boson via
\beq
\MW^2 \equiv (1+ \Delta x) M^2_{W^0} \equiv xM^2_{W^0},
\label{7}
\eeq
while the {\it ``coupling parameter''}  
$\Delta y$ specifies SU(2) violation among the $W^\pm$ and $W^0$ couplings 
to fermions,
\beq
g^2_{W^\pm} (0) \equiv M^2_{W^\pm} 4 \sqrt 2 G_\mu = (1 + \Delta y) g^2_{W^0}
(\MZ^2) \equiv y g^2_{W^0} (\MZ^2).
\label{8}
\eeq
{}Finally, the {\it ``mixing parameter''} $\varepsilon$ refers to 
the mixing strength in the neutral vector boson sector and quantifies the
deviation 
of $\swbar^2$ from $e^2 (\MZ^2)/g^2_{W^0} (\MZ^2)$,
\beq
\swbar^2 \equiv \frac{e^2 (\MZ^2)}{g^2_{W^0} (\MZ^2)} (1 - \varepsilon ),
\label{9}
\eeq
thus allowing for an unconstrained mixing strength\ucite{kn91,HS} in the neutral
vector-boson sector. 
The effective Lagrangian incorporating the mentioned sources of SU(2)
violation for $W$ and $Z$ interactions with leptons is given by\ucite{DKK,BKK}
\beq
{\cal L}_C = -\frac{1}{2} W^{+\mu\nu}W^-_{\mu\nu}
+ \frac{g_\PWpm}{\sqrt 2} \left( j^+_\mu W^{+\mu} + h.c.\right)
+ \MWpm^2 W^+_\mu W^{-\mu}
\label{1a}
\eeq
and 
\beqar
{\cal L}_N & = & - \frac{1}{4} Z_{\mu\nu} Z^{\mu\nu} +
\frac{1}{2} \frac{\MWO^2}{1-\bar\sw^2(1-\varepsilon)}Z_{\mu}Z^{\mu}
- \frac{1}{4} A_{\mu\nu} A^{\mu\nu} \nn\\
&& - e j_{em}^\mu A_\mu +
\frac{g_\PWO}{\sqrt{1-\bar\sw^2 (1-\varepsilon)}}
\left( j^\mu_3 - \bar\sw^2 j^\mu_{em}\right) Z_\mu.
\label{10a}
\eeqar

{}For the observables $\swbar^2 , \MW$ and $\Gl$,
from \refeq{1a} and \refeq{10a} one obtains
\begin{eqnarray}
\swbar^2 (1 - \swbar^2 ) & =& {{\pi \alpha (\MZ^2)}\over{\sqrt 2 G_\mu \MZ^2}}
{y\over x} (1 - \varepsilon) {1\over{\left( 1 + {\swbar^2\over{1 - \swbar^2}}
\varepsilon \right)}}, \nn  \\
{\MW^2\over \MZ^2} & =& (1 - \swbar^2 ) x \left( 1 + {\swbar^2 \over
{1 - \swbar^2}} \varepsilon \right) , \nn\\
\Gl & =& {{G_\mu M^3_Z}\over{24\pi \sqrt 2}} \left( 1 + (1 - 4 \swbar^2 )^2
\right) {x\over y} \left( 1 - {{3\alpha}\over{4\pi}} \right). 
\label{11}
\end{eqnarray}
{}For $x = y = 1$ (i.e., $\Delta x = \Delta y = 0$) and $\varepsilon = 0$ one
recovers the $\alpha (\MZ^2)$-Born approximation,
$\swbar^2 = s^2_0$, discussed previously. 

The extension\ucite{DKS} of the effective Lagrangian \refeq{10a} 
to interactions of neutrinos and
quarks requires the additional coupling parameters $\Delta y_\nu$ for the
neutrino, $\Delta y_b$ for the bottom quark, and $\Delta y_h$ for the remaining
light quarks. In the analysis of the data, for $\Delta y_\nu$ and $\Delta y_h$
which do not involve the non-Abelian structure of the theory, the SM
theoretical results may be inserted without loss of generality
as far as the guiding principle of separating vector-boson--fermion
interactions from interactions containing non-Abelian couplings is
concerned.

We note that the parameters in our effective Lagrangian are related\ucite{DKS} to the 
parameters $\varepsilon_{1,2,3}$ and $\epsilon_b$, 
introduced\ucite{al93} by isolating the
quadratic $\Mt$ dependence, 
\beqar
\parbox{6cm}{$\eps_1=\De x-\De y+0.2\times 10^{-3},$} &&
\eps_2=-\De y+0.1\times 10^{-3}, \nn\\
\parbox{6cm}{$\eps_3=-\eps+0.2\times 10^{-3},$} &&
\eps_\Pb=-\De\yb/2-0.1\times 10^{-3}.
\label{10b}
\eeqar
Essentially the two sets of parameters only differ in $\varepsilon_1$. As
$\varepsilon_1$ contains a linear combination of $\Delta x$ 
and $\Delta y$, the 
$\MH$-dependent bosonic corrections in $\Delta x$ are confused with the 
$\MH$-insensitive bosonic corrections in $\Delta y$, i.e. with our choice
of parameters the $M_H$-insensitive corrections are isolated and appear
in the single parameter $\Delta y$ only.
The theoretically interesting, but numerically irrelevant additive terms in
\refeq{10b},
considerably smaller than $1 \times 10^{-3}$, originate from a refinement in the
mixing involved in Lagrangian \refeq{10a} and a corresponding refinement in
\refeq{11}. We refer to the original paper\ucite{DKS} for details. 
 
By linearizing the equations in \refeq{11} 
with respect to $\Delta x, \Delta y$ and 
$\varepsilon$ and  
inverting them, $\Delta x , \Delta y$ and $\varepsilon$ may 
be deduced from the experimental data on $\swbar^2, \Gl$ and 
$\MW$.
Inclusion of the hadronic $Z$ observables requires that $\Delta x , \Delta y ,
\varepsilon$ and $\Delta y_b$ are fitted to the experimental data. 
Actually, one
finds that the
results for $\Delta x , \Delta y , \varepsilon$ are hardly affected by inclusion of
the hadronic observables.  
On the other hand, $\Delta x , \Delta y ,\varepsilon$ and $\Delta y_b$ may be
theoretically determined 
in the standard electroweak theory 
at the one-loop level, strictly discriminating 
between pure fermion-loop predictions and the rest which contains the unknown 
bosonic couplings. The most recent 1996 update\ucite{DS} of such an 
analysis\ucite{DSW,DKK,DKS} is shown in fig.~2.
\begin{figure}
\begin{center}
\begin{picture}(15,14.5)
\put(-2.0,-4) {\includegraphics{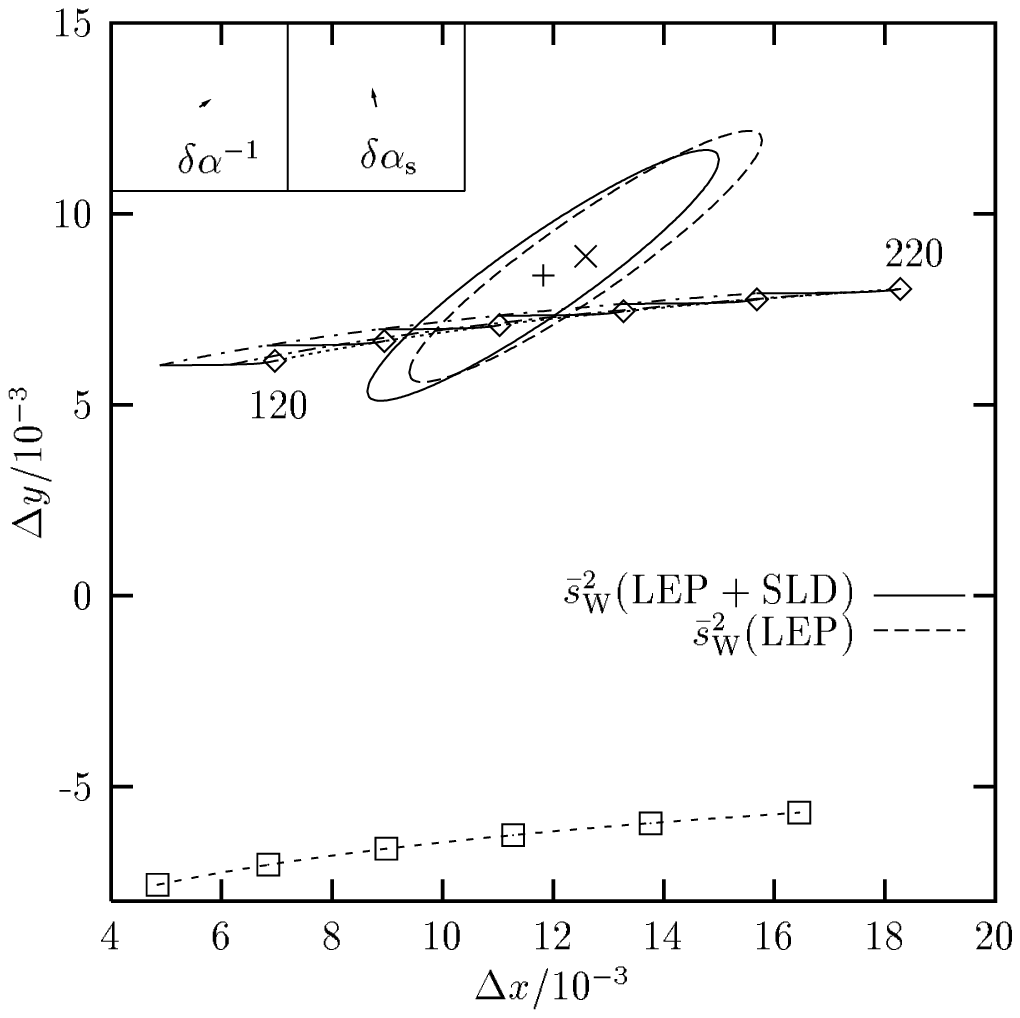}}
\put( 4.7,-4) {\includegraphics{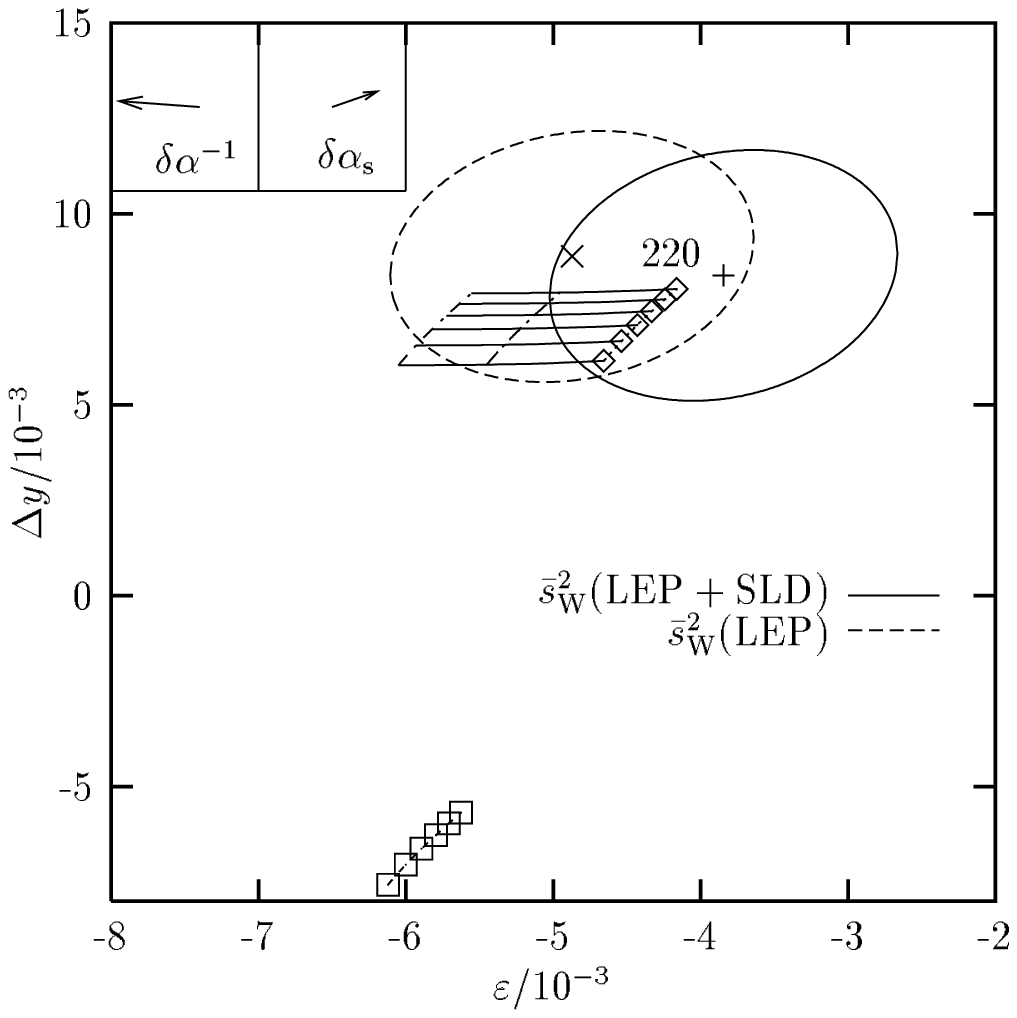}}
\put(-2.0,-10.5){\includegraphics{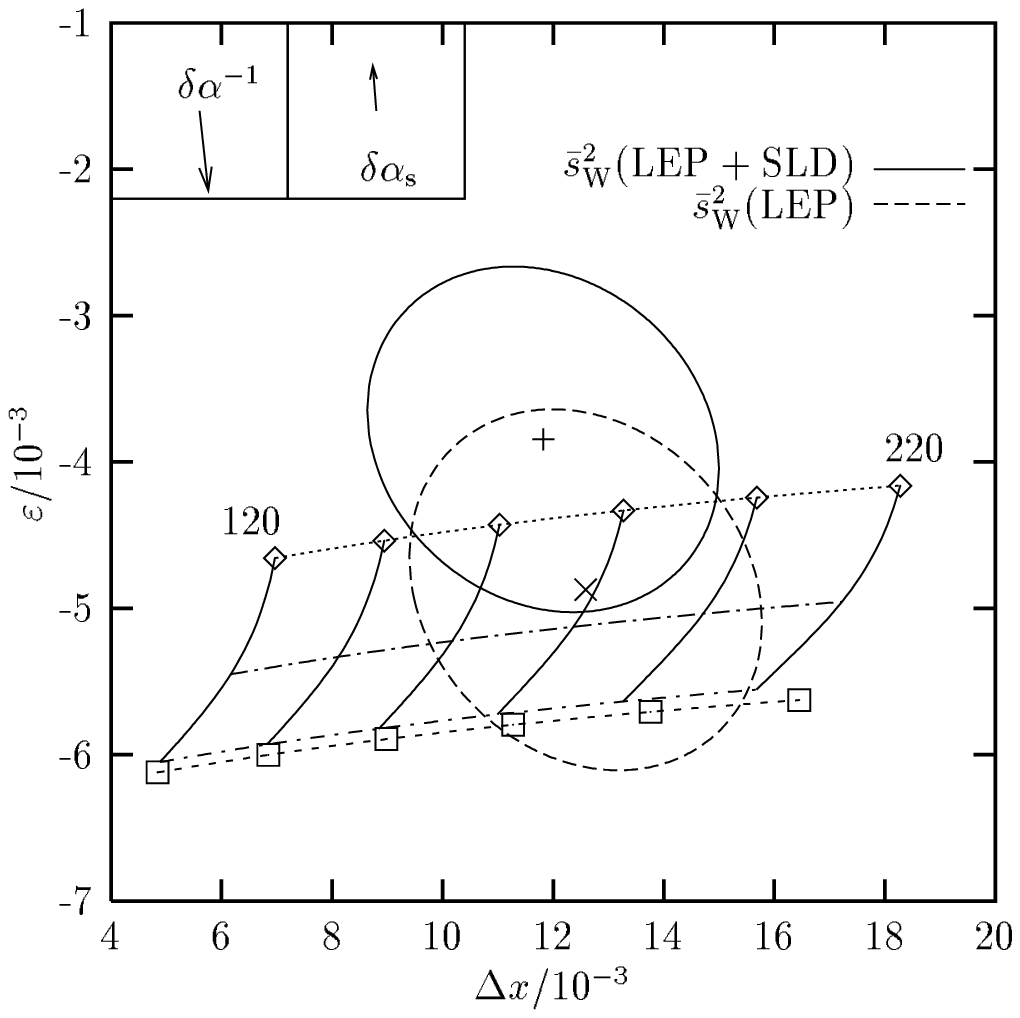}}
\put( 4.7,-10.5){\includegraphics{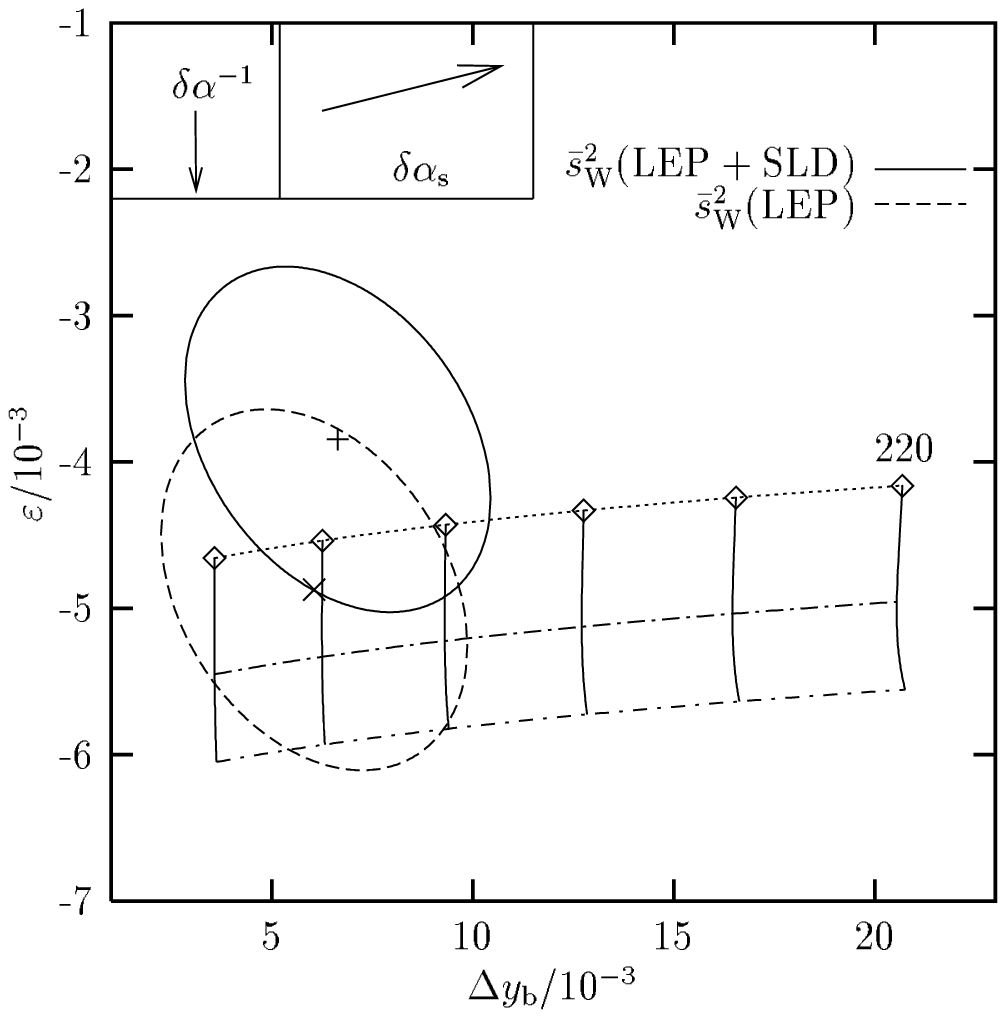}}
\end{picture}
\end{center}
\caption{\it
The projections of the $1\si$ ellipsoid of 
the electroweak parameters $\De x$, $\De y$, $\eps$,
$\De\yb$ obtained from the 1996 set of data in comparison
with the SM predictions. Both the
results obtained from using $\swbar^2(\LEP)$ and $\swbar^2(\LEP+\SLD)$
as experimental input are shown. The full SM predictions
correspond to Higgs-boson masses of $100\GeV$ (dotted with diamonds), 
$300\GeV$ (long-dashed dotted) and $1\TeV$ (short-dashed dotted)
parametrized by the top-quark mass ranging from
$120\GeV$ to $220\GeV$ in steps of $20\GeV$. 
The pure fermion-loop
prediction is also shown (short-dashed curve with squares) for the
same values of $\Mt$.
The arrows indicate the shifts of the centres of the ellipses upon
changing $\alpz^{-1}$ to $\alpz^{-1}+\de\alpz^{-1}$ and $\alpsz$ to
$\alpsz+\de\alpsz$. (From \citere{DS})}
\label{fig:xyeb}
\efi

According to fig.~2, the data in the $(\varepsilon , \Delta x)$ plane are  
consistent with the SM predictions obtained by approximating $\Delta x$
and $\varepsilon$ by their 
pure fermion-loop values, 
\begin{eqnarray}
\Delta x & =& \Delta x_{{\rm ferm}} (\alpha (\MZ^2) , s^2_0 , \Mt^2 \ln \Mt ) 
+ \Delta x_{{\rm bos}} (\alpha (\MZ^2) , s^2_0 , \ln \MH^2) \nn \\
  & \cong & \Delta x_{{\rm ferm}} (\alpha (\MZ^2) , s^2_0 , \Mt^2 ,\ln \Mt) ,  
\nn \\[.3em]
\varepsilon & = & \varepsilon_{{\rm ferm}} (\alpha (\MZ^2) , s^2_0 , \ln \Mt )+ 
\varepsilon_{{\rm bos}} (\alpha (\MZ^2) , s^2_0 , \ln \MH^2 ) \nn\\
  &\cong & \varepsilon_{{\rm ferm}} (\alpha (\MZ^2) , s^2_0 , \ln \Mt ) . 
\label{12}
\end{eqnarray}
The small contributions of 
$\Delta x_{{\rm bos}}$ and $\varepsilon_{{\rm bos}}$ to $\Delta x$ and $\varepsilon$,
respectively, and the logarithmic dependence on the Higgs mass, $\MH$,
imply the well-known result 
that the data are fairly insensitive to the mass of the Higgs scalar. 
It is instructive to also note the numerical results for $\Delta x_{{\rm ferm}}$
and $\varepsilon_{{\rm ferm}}$, obtained in the Standard Model. They are given by\ucite{DKK}
\beqar
\Delta x_{{\rm ferm}} & = & (2.61 t + 1.34 \log (t) + 0.52) \times 10^{-3} , \nn\\
\varepsilon_{{\rm ferm}} & = & (- 0.45 \log (t) - 6.43) \times 10^{-3} ,
\label{x}
\eeqar
with $t \equiv \Mt^2 / \MZ^2$. The mass parameter $\Delta x$ is dominated by the
$\Mt^2$ term\ucite{VELT} due to weak isospin breaking induced by the top quark, while
$\varepsilon$ is dominated by the constant term due to mixing 
among the neutral vector bosons induced by the light
leptons and quarks. 
 
In distinction from the results for $\Delta x$ and $\varepsilon$, 
where the fermion loops by themselves are consistent with the data, 
a striking effect appears in the plots showing $\Delta y$. 
The predictions are
clearly inconsistent with the data, unless the fermion-loop contributions to $\Delta y$ 
(denoted by lines with small squares) are supplemented by an 
additional term, which in the standard electroweak theory is due to bosonic effects, 
\beq
\Delta y = \Delta y_{{\rm ferm}} (\alpha (\MZ^2) , s^2_0 , \ln \Mt ) + \Delta y_{{\rm 
bos}} (\alpha (\MZ^2) , s^2_0). 
\label{13}
\eeq
Remembering that $\Delta y$, according to \refeq{8}, 
relates the coupling of the $W^\pm$ boson to leptons as measured in 
$\mu^\pm$ decay, to the coupling of the neutral member, $W^0$, 
of the vector-boson triplet at the scale $\MZ$, 
it is not surprising that $\Delta y_{\bos}$ contains vertex and box corrections
originating from $\mu^\pm$ decay as well as vertex corrections at the $W^0
f\bar f$ ($Z f \bar f$) vertex.
While $\Delta y_{{\rm bos}}$ obviously depends on the trilinear couplings among the vector
bosons, it is insensitive to $\MH$. 
The experimental data have accordingly become accurate enough to  
isolate loop effects which are insensitive to $\MH$, but depend on the 
self-interactions of the vector bosons, in particular on the trilinear
non-Abelian couplings 
entering the $W f\bar f^\prime$ and $W^0 f\bar f$ ($Z f \bar f)$ 
vertex corrections.

With respect to the interpretation of the coupling parameter, 
$\Delta y$, one further step\ucite{DSW} may appropriately be taken. Introducing the 
coupling of the $W$ boson to leptons, $g_{W^\pm} (\MW^2)$, as defined by the
leptonic $W$-boson width, 
in addition to the previously used low-energy coupling, $g_{W^\pm} (0)$, defined
by the Fermi constant in \refeq{8},
\beq 
\Gamma^W_l = g^2_{W^\pm} (\MW^2) \frac{\MW}{48\pi} \left( 1 + c^2_0
\frac{3\alpha}{4\pi} \right),
\label{14}
\eeq
the coupling parameter, $\Delta y$, in linear 
approximation may be split
into two additive terms, 
\beq
\Delta y = \Delta y^{\SC} + \Delta y^{\IB}.
\label{15}
\eeq
While $\Delta y^{\SC}$ (where ``$\SC$'' stands for ``scale change'') furnishes the
transition from $g_{W^\pm} (0)$ to $g_{W^\pm} (\MW^2)$, 
\beq
g^2_{W^\pm} (0) = (1 + \Delta y^{\SC}) g^2_{W^\pm} (\MW^2),
\label{16}
\eeq
the parameter $\Delta y^{\IB}$ (where ``$\IB$'' stands for ``isospin breaking'')
relates the charged-current and neutral current couplings at the high-mass
scale $\MW\sim\MZ$, 
\beq
g^2_{W^\pm} (\MW^2) = ( 1 + \Delta y^{\IB}) g^2_{W^0} (\MZ^2),
\label{17}
\eeq
to each other. Note that $\Delta y^{\SC}$ according to \refeq{14} with \refeq{16}
and \refeq{8} can be uniquely extracted from the observables $\MW , \Gamma^W_l$
together with $G_\mu$. 
\begin{figure}
\begin{center}
\begin{picture}(15,8.5)
\put(-1.0,-9.7){\includegraphics{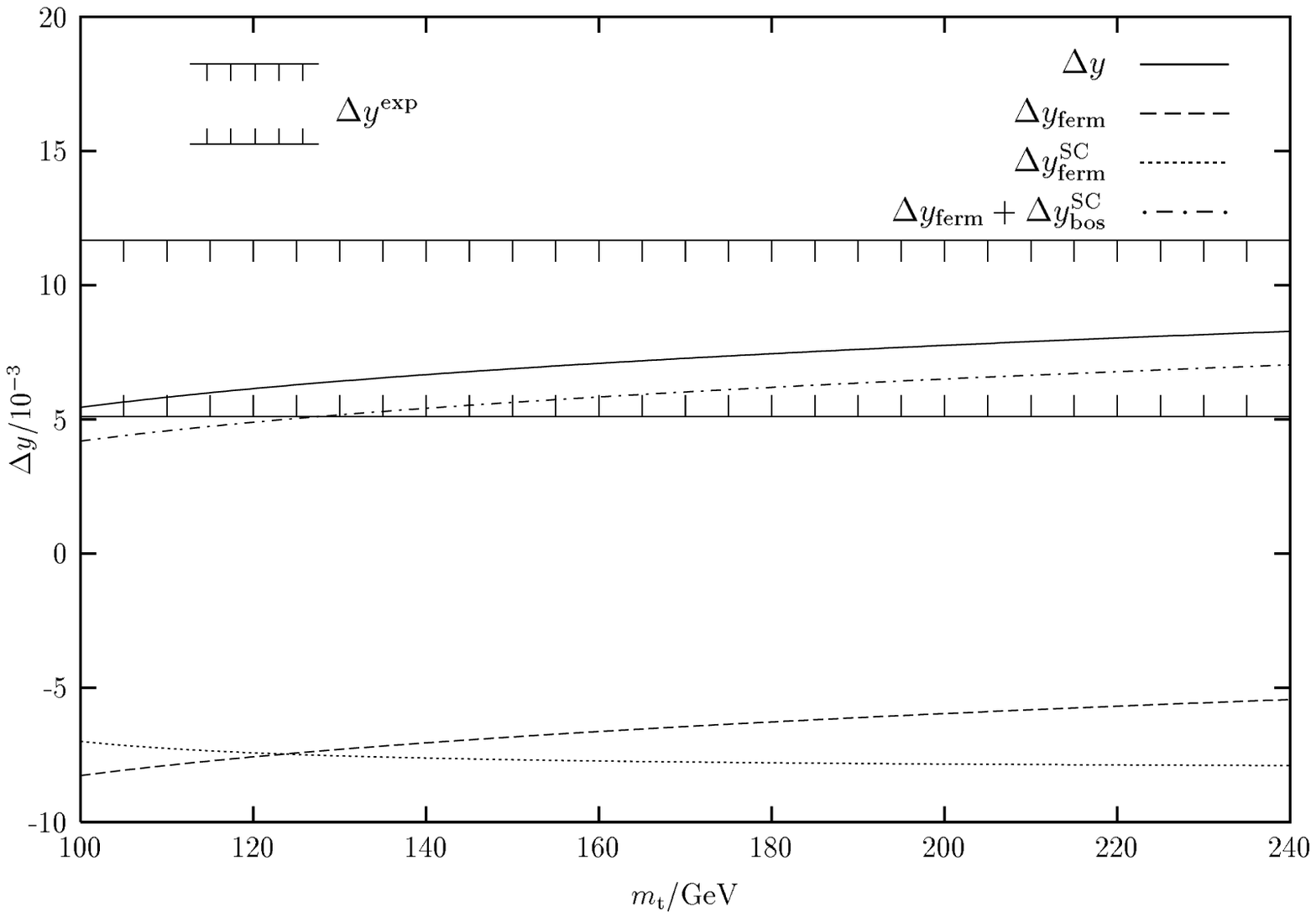}}
\end{picture}
\end{center}
\caption{\it The one-loop SM predictions for $\De y$, $\De
y_{\fer}$, $\De y^{\SC}_{\fer}$, and $(\De y_{\mathrm
ferm} + \De y^{\SC}_{\bos})$ as a function of $m_\Pt$.
The difference between the curves for $\De y$ and $(\De y_{\fer} +
\De y^{\SC}_{\bos})$ corresponds to the small contribution of $\De
y^{\IB}_{\bos}$.
The experimental value of $\Delta y$,
$\Delta y^{\mathrm exp} = (8.4 \pm 3.3) \times 10^{-3}$, is indicated
by the error band (From \citere{DSW}, 1996 update).}
\label{fig:delysc}
\efi

\btab
\bce
\begin{tabular}{|l|c|c|c|} \hline
~ & $\Delta y_{{\rm ferm}} \times 10^3$ & $\Delta y_{{\rm bos}} \times 10^3$ & 
$\Delta y \times 10^3$ \\   \hline
SC & $-7.8$ & 12.4 & 4.6  \\ \hline
IB ($\Mt = 175$ GeV) & 1.5 & 1.2 & 2.7  \\   \hline
SC + IB & $-6.3$ & 13.6 & 7.3  \\  \hline  
\end{tabular}
\vspace*{-.4cm}
\ece
\caption[]{\it The different contributions (see \refeq{15})
to the coupling parameter $\Delta y$
(from \citere{DSW}).} 
\etab   

As seen in tab.~2 and fig.~3, the fermion-loop and the bosonic contributions to $\Delta y$ are 
opposite in sign, and both are dominated by their scale-change parts, $\Delta
y^{\SC}$. 
Once, $\Delta y^{\SC}_{{\rm bos}}$ is taken into account, practically no further bosonic 
contributions are needed. 
 
The bosonic loops necessary for agreement
with the data are accordingly recognized as charged-current corrections related
to the use of the low-energy parameter $G_\mu$ in the analysis of the data at
the Z~scale.
Their contribution, due to a gauge-invariant combination of vertex, box and
vacuum-polarization, is opposite in sign and somewhat larger than the
contribution due to fermion-loop vacuum polarization, the increase in
$g_{W^\pm}$ due to fermion loops thus becoming overcompensated by bosonic
corrections.

We note that the coupling $g_{W^\pm} (M^2_W)$, obtained from $G_\mu, M_W$
and $\Delta y^{SC}$, is most appropriate to define an (improved) Born
approximation\ucite{Greece} for $e^+e^- \to W^+W^-$ at LEP2 energies.

Once the input parameters at the Z~scale, $\MZ$ and $\alpha (\MZ^2)$, are
supplemented by the coupling $g_{W^\pm} (\MW^2)$, also defined at this
scale and replacing $G_\mu$, all relevant radiative
corrections are 
contained in $\De x_{\fer}$, $\eps_{\fer}$, and $\De\yb$, and are
either related to weak isospin breaking by the top quark or due
to mixing effects induced by the light leptons and quarks and the top
quark. Compare the numerical results for $\Delta x_{{\rm ferm}}$ and
$\varepsilon_{{\rm ferm}}$ in \refeq{x}.
In addition to $\Delta x_{\rm ferm}$ and $\varepsilon_{\rm ferm}$, 
there is a (small) $\log (\Mt)$ 
isospin-breaking contribution to $\Delta y$ as shown in tab.~2, 
and an even smaller bosonic isospin-breaking contribution. 

In fig.~2, we also show the result for $\Delta y_b$ in the $(\Delta y_b ,
\varepsilon)$ plane. The SM prediction for $\Delta y_b$, as a
consequence of a quadratic dependence on $\Mt$, is similar in magnitude to the
one for $\Delta x$. The experimental result for $\Delta y_b$ at the $1\sigma$
level almost includes the theoretical expectation implied by the Tevatron
measurement of $\Mt^{{\rm exp}} = 175 + 6$ GeV. This reflects the fact that the
1996 value of $\Rb$ from tab.~1 is approximately consistent with theory, since
the $\Rb$ enhancement, present in the 1995 data\ucite{LEPEWWG9502} 
has practically gone
away. I will come back to this point when discussing the bounds on $\MH$ implied
by the data. 
 
\subsection{Empirical Evidence for the Higgs Mechanism?}

As the experimental results for $\Delta x$ and $\varepsilon$ are well represented by
neglecting all effects with the exception of fermion loops, and as the bosonic 
contribution to $\Delta y$, which is seen in the data, 
is independent of $\MH$, the 
question as to the role of the Higgs mass and the concept of the Higgs 
mechanism\ucite{HK} with 
respect to precision tests immediately arises.
 
More specifically, one may ask the question whether the experimental results
(i.e.\ $\Delta x, \Delta y , \varepsilon$, $\De\yb$)
can be predicted even without the very concept of the 
Higgs mechanism. 
 
In \citere{DGS} we start from the well-known fact that the standard electroweak 
theory without Higgs particle may credibly be reconstructed\ucite{HS}
within the framework of a 
massive vector-boson theory (MVB) 
with the most general mass-mixing term which preserves 
electromagnetic gauge invariance. This theory is then cast into a form which is 
invariant under local SU(2)$\times$U(1) 
transformations by introducing three auxiliary
scalar fields \'a la Stueckelberg\ucite{ST,KG}. 
As a consequence, loop calculations may be carried out
in an arbitrary $R_\xi$ gauge
in close analogy to the SM, even though the non-linear realization of
the SU(2)$\times$U(1) symmetry, obviously, does not imply renormalizability of
the theory.
 
Explicit loop calculations show that indeed the Higgs-less observable $\Delta y$, 
evaluated in the MVB, coincides with $\Delta y$ evaluated in 
the standard electroweak theory, i.e.\ in particular for the bosonic part,
we have\footnote{Actually, in the SM there is an additional contribution
of ${\cal O}(1/\MH^2)$ which is irrelevant numerically for 
$\MH \gsim 100$ GeV.
Note that the $\MH$-dependent contributions to interactions violating
custodial SU(2) symmetry turn out to be suppressed\ucite{he94sdcgk} by a
power of $1/\MH^2$ in the SM relative to the expectation from
dimensional analysis.
The absence of a $\log\MH$ term in $\De y$ and the absence of
a $\MH^2\log\MH$ term in $\De x$ in the SM thus appear on equal footing 
from the point of view of custodial SU(2) symmetry. 
In contrast,
no suppression relative to dimensional analysis is present in the 
mixing parameter $\varepsilon$, which does not violate custodial SU(2) 
symmetry.
}
\beq
\Delta y_{{\rm bos}}^{{\rm MVB}} \equiv \Delta y_{{\rm bos}}^{{\rm SM}}.
\label{18}
\eeq
In the case of  $\Delta x_{{\rm bos}}$ and $\varepsilon_{{\rm bos}}$, one finds that the 
MVB and the SM differ by the replacement
$\ln \MH \Leftrightarrow \ln \Lambda $ ,
where $\Lambda$ denotes an ultraviolet cut-off.
{}For $\Lambda \lsim 1$ TeV, accordingly, 
\beq
\Delta x^{{\rm MVB}} \cong \Delta x_{{\rm ferm}}^{{\rm MVB}} =
\Delta x_{{\rm ferm}}^{{\rm SM}}, \qquad
\varepsilon^{{\rm MVB}} \cong \varepsilon_{{\rm ferm}}^{{\rm MVB}} =
\varepsilon_{{\rm ferm}}^
{{\rm SM}}.
\label{19}
\eeq
 
In conclusion, the MVB can indeed be evaluated at one-loop level
at the expense of introducing a logarithmic cut-off, $\Lambda$. This cut-off only 
affects the mass parameter,
$\Delta x$, and the mixing parameter,
$\varepsilon$, whose bosonic contributions cannot be well resolved 
experimentally anyway. 
 
The quantity $\Delta y$, whose bosonic contributions are essential for agreement with 
experiment, is independent of the Higgs mechanism, i.e. it is convergent for
$\Lambda \rightarrow \infty$ in the MVB theory.  
It depends on the non-Abelian couplings
of the vector bosons among each other, which enter the vertex corrections at the 
$W$ and $Z$ vertices. Even though the data cannot discriminate between the 
MVB and the SM with Higgs scalar, the Higgs mechanism 
nevertheless yields 
the only known simple physical realization of the cut-off $\Lambda$
(by $\MH$) which guarantees renormalizability. 
 
\subsection{Bounds on the Higgs-Boson Mass}

We return to the description of the data in the SM, and
in particular discuss the question, in how far the mass of the Higgs boson
can be deduced from the precision data. 

In Section 1.3 we noted that the full (logarithmic) dependence on $\MH$ is
contained in the mass parameter, $\Delta x$, and in the mixing parameter,
$\varepsilon$. The experimental restrictions on $\MH$ may accordingly be visualized
by showing the contour of the data in the $(\Delta x, \varepsilon )$ plane for the
fixed (theoretical) value of $\Delta y \cong 7\times 10^{-3}$ (corresponding to $\Mt =
175 \pm 6$ GeV) in comparison with the $\MH$-dependent SM predictions for
$\Delta x$ and $\varepsilon$. Figure 4 illustrates the delicate dependence of 
bounds
for $\MH$ on the experimental input for $\swbar^2, \alpha (\MZ^2)$ and
$\Mt^{{\rm exp}}$. The bounds on $\MH$, one can read off from fig.~4, are
qualitatively in agreement of the results of the fits to be discussed next.  
\begin{figure}
\begin{center}
\begin{picture}(15,7)
\put( 1.8,6.1){$\swbar^2(\LEP+\SLD)$}
\put(9.1,6.1){$\swbar^2(\LEP)$}
\put(-2.0,-10.4){\includegraphics{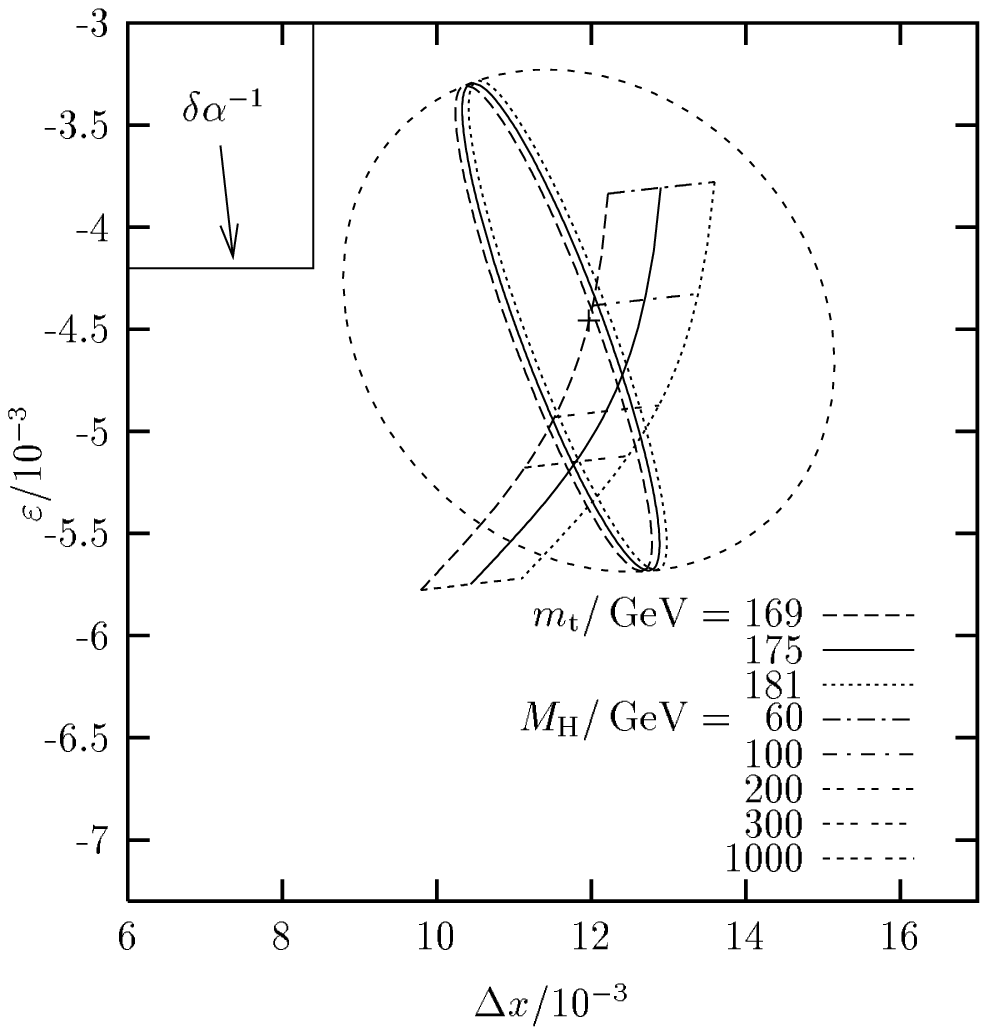}}
\put( 4.7,-10.4){\includegraphics{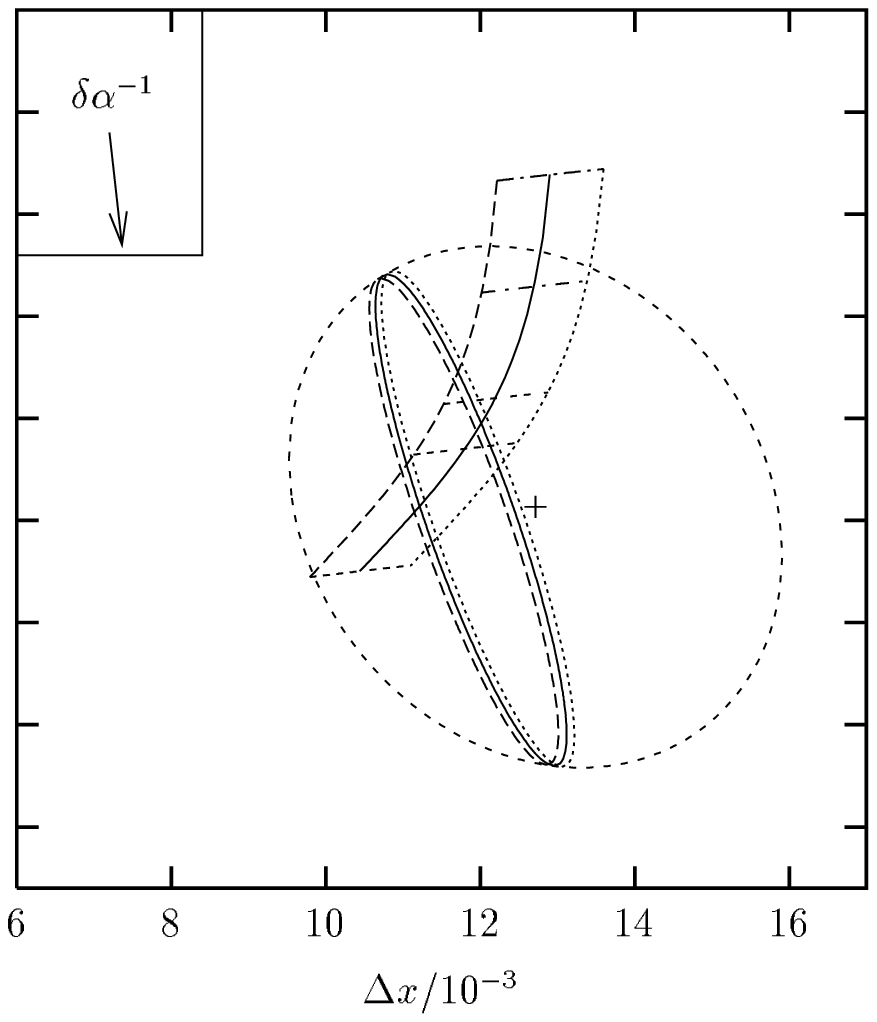}}
\end{picture}
\end{center}
\caption{\it The
$1\sigma$ contour of the experimental data in the
$(\Delta x,\varepsilon)$
plane defined by $\Delta y \protect\cong 7 \times 10^{-3}$
(corresponding to $\Mt = 175 \pm 6$ GeV). The cut of the contour with the
SM predictions for $\Mt = 175 \pm 6$ GeV yields the experimental
bounds on $\MH$. The projection of the data ellipsoid on the 
$(\Delta x,\varepsilon)$ plane, also shown, differs slightly from the one 
in fig.~2,
since the data from the leptonic sector only were used for the present figure.
}
\efi

Precise bounds on $\MH$ require a fit to the experimental data. In order to
account for the experimental uncertainties in the input parameters of $\alpha
(\MZ^2), \alps (\MZ^2)$ and $\Mt$, 
four-parameter $(\Mt,
\MH, \alpha (\MZ^2), \alps (\MZ^2))$ fits to various sets of observables from
tab.~1 were actually performed in Refs.\cite{DSCW,DS}. 
$\MH$ and $\alps (\MZ^2)$ were treated as free fit
parameters, while for $\alpha (\MZ^2)$ and $\Mt$ 
the experimental constraints from tab.~1 were used.

The results of the 1996 update (taken from \citere{DS}) of the 
fits\ucite{DSCW}\footnote{Compare also \citere{LEPEWWG9602,ho96} for 
$\MH$-fits to the 1996
electroweak data, and \citere{mo94,el95} for $\MH$ fits to previous sets 
of data.}
are presented in the plots of
$\Delta \chi^2 \equiv \chi^2 - \chi^2_{\min}$ against $\MH$ of fig.~5.
\begin{figure}
\begin{center}
\begin{picture}(15,11.5)
\put(-2,-9.8){\includegraphics{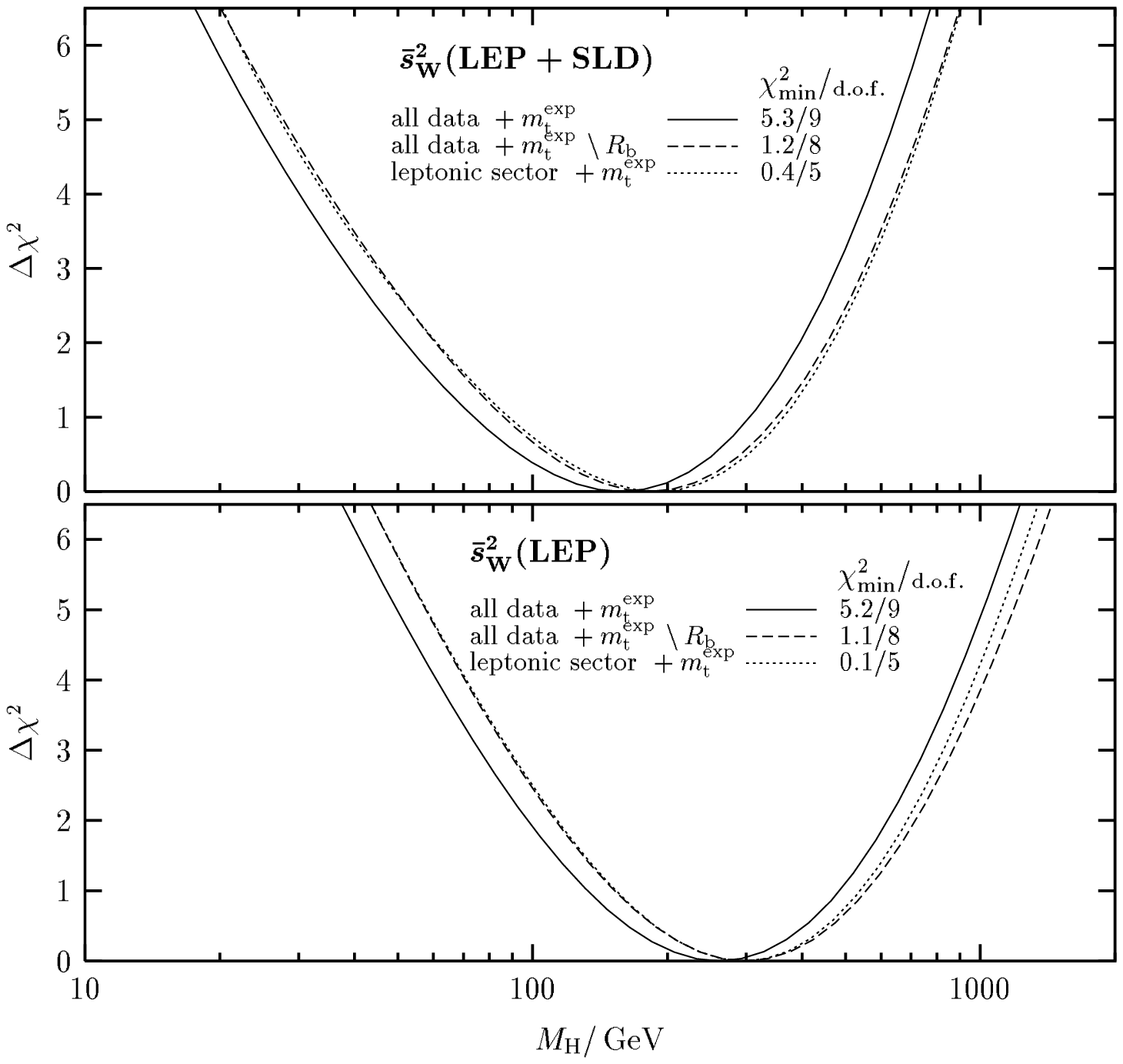}}
\end{picture}
\end{center}
\caption{\it $\De\chi^2=\chi^2-\chi^2_{\mathrm{min}}$ is plotted against
$\MH$ for the $(\Mt,\MH,\alpz,\alpsz)$ fit to various sets of 
observables.
{}For a chosen input for
$\swbar^2$, as indicated, we show the result of a fit to
\protect\\
(i) the full set of 1996 data, 
$\swbar^2$, $\MW$, $\GT$, $\si_{\mathrm{h}}$, $R$, $\Rb$, $\Rc$, 
together with
$\Mt^{\exp}$, $\alpz$,
\protect\\
(ii) the 1996 set of (i) upon exclusion of $\Rb$,
\protect\\
(iii) the 1996 ``leptonic sector'' of $\swbar^2$, $\MW$, $\Gl$, 
together with $\Mt^{\exp}$, $\alpz$. (From \citere{DS})
}
\label{fig:Dchi}
\efi
\begin{figure}
\begin{center}
\begin{picture}(15.1,15.5)
\put(0.8,-0.2){\includegraphics{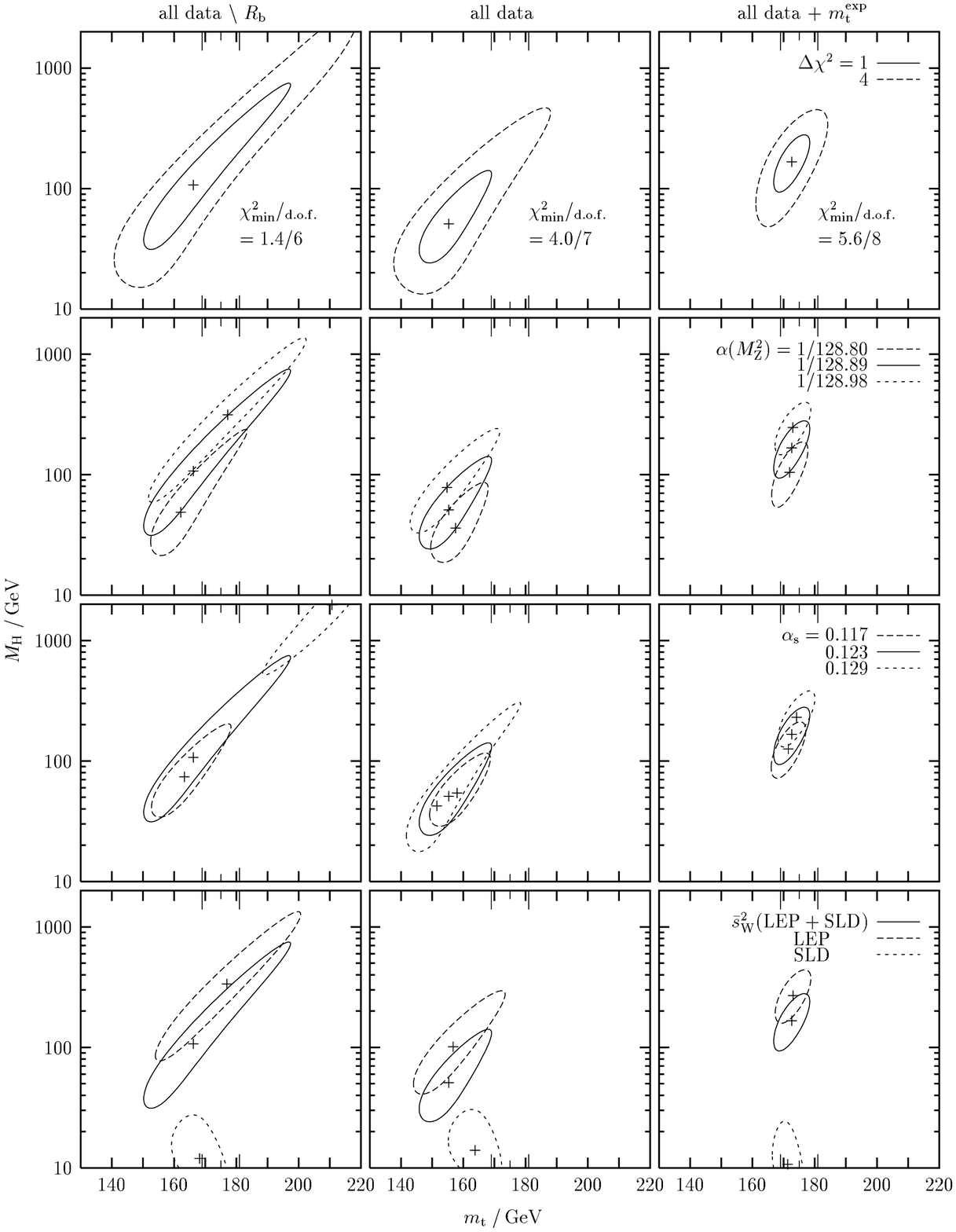}}
\end{picture}
\end{center}
\caption{\it
The results of the two-parameter $(\Mt,\MH)$ fits within the SM
are displayed in the $(\Mt, \MH)$ plane. The different columns refer 
to the sets of experimental data used in the corresponding 
fits, \protect\\
(i) ``all data $\backslash\Rb$'': $\swbar^2(\LEP+\SLD)$, 
$\MW$, $\GT$, $\si_{\mathrm{h}}$, $R$, $\Rc$, 
\protect\\
(ii) ``all data'': $\Rb$ is added to set (i), 
\protect\\
(iii) ``all data + $\Mt^{\exp}$'': $\Rb,\Mt^{\exp}$ are added to the set (i).
\protect\\
The second and third row shows the shift resulting from changing 
$\alpz^{-1}$ and $\alpsz$, respectively,
by $1\sigma$ in the SM prediction.
The fourth row shows the effect of replacing $\swbar^2(\LEP+\SLD)$ by 
$\swbar^2(\LEP)$ and $\swbar^2(\SLD)$ in the fits.
Note that the $1\sigma$ boundaries given in the first row 
are repeated identically in each row, in order to facilitate comparison with 
other boundaries. 
The value of $\chidof$ given in the plots refers to the 
central values of $\alpz^{-1}$ and $\alpsz$.
In all plots the empirical value 
of $\Mt^{\exp} = 175 \pm 6 \GeV$ is also indicated. (From \citere{DS})}
\label{fig:mtmhfit}
\efi

As $\chi^2_{\min}$ is smallest for the fit to the ``leptonic sector'' of $\bar
s^2_W , \MW , \Gl$ together with $\Mt^{\exp}$, and $\alpha (\MZ^2)$, while the
$1\sigma$ errors are approximately the same in the three fits shown in fig.~5,
we quote the result from the leptonic sector as the most reliable one,
\beqar
\MH = 190^{+174}_{-102} {\rm GeV}, & \quad {\rm using} \quad 
& \swbar^2 ({\rm LEP +
  SLD})_{'96} = 0.23165 \pm 0.00024, \nn  \\
\MH = 296^{+243}_{-143} {\rm GeV}, & \quad {\rm using} \quad 
& \swbar^2 ({\rm LEP})_{'96}
= 0.23200 \pm 0.00027  
\label{20}
\eeqar
based on the 1996 set of data. It implies the $1\sigma$ bounds of 
$\MH \lsim 360$ GeV and $\MH \lsim 540$ GeV,
using $\swbar^2$ (LEP + SLD) and $\swbar^2$ (LEP), respectively, and 
\beqar
\MH \lsim 550 {\rm GeV}~(95\% {\rm C.L.}) & \quad {\rm using} \quad 
& \swbar^2 ({\rm SEP +
  SLD})_{'96}, \nn \\
\MH \lsim 800 {\rm GeV}~(95\% {\rm C.L.}) & \quad {\rm using} \quad 
& \swbar^2 ({\rm LEP})_{'96}.
\label{21}
\eeqar
The fact that the results \refeq{20} and \refeq{21} do not require $\alps
(\MZ^2)$ as input parameter (apart from two-loop effects), and accordingly are
independent of the uncertainties in $\alps (\MZ^2)$, provides an additional reason
for the restriction to the leptonic sector when deriving bounds for
$\MH$. Moreover, we note that according to fig.~5 the results for $\MH$ given by
\refeq{20} and \refeq{21} practically
do not change if the $\alps (\MZ^2)$-dependent
observables, $\GT$ and $\Gh$, the total and hadronic $Z$ widths, are
included in the fit. Inclusion of $\GT$ and $\Gh$ 
provides important information on
$\alps (\MZ^2)$, however. One obtains\ucite{DS} 
$\alps (\MZ^2) = 0.121 \pm 0.003$
and $\alps (\MZ^2) = 0.123 \pm 0,003$ depending on whether $\swbar^2$ (LEP
+ SLD) or $\swbar^2$ (LEP) was used in the fit. Both values are consistent
with the event-shape result given in tab.~1. The impact of also including $\Rb$
in the fit, also shown in fig.~5, will be commented upon below. Inclusion or
exclusion of $\Rc$ is unimportant, as the error in $\Rc$ is considerable. 

As mentioned, the above results on $\MH$ are based on the 1996 set of data 
[4,5,6] which
was presented at the Warsaw International Conference on High Energy Physics 
which took place towards the end of July 1996. 
Two results presented in Warsaw are of particular importance
with respect to the bounds on $\MH$. 

{}First of all, the value of $\Mt = 175 \pm 6$ GeV reported in Warsaw and given in
tab.~1 
is significantly more precise than the 1995 result\ucite{LEPEWWG9502} of $\Mt =
180 \pm 12$ GeV. The decrease in the error on $\Mt$, due to the $(\Mt, \MH)$
correlation in the SM predictions for the observables, 
clearly visible in fig.~4, led to a
substantially narrower $\Delta\chi^2$ distribution in fig.~5 compared with the
results based on the 1995 set of data. Indeed, the 1995 leptonic set of
data had implied\ucite{DSCW}
\beqar
\MH & = & 152^{+282}_{-106}\GeV \quad {\rm using} \quad \swbar^2~({\rm
LEP + SLD})_{'95} = 0.23143\pm 0.00028, \nn \\
\MH & = & 353^{+540}_{-224}\GeV \quad {\rm using} \quad \swbar^2~({\rm LEP})_{'95} =
0.23186 \pm 0.0034 , 
\label{22}
\eeqar
i.e., central values similar to the ones in \refeq{20}, but with substantially larger
errors. 

The second and most pronounced change occurred in the result for $\Rb \equiv
\Gamma_b / \Gh$. The enhancement in the 1995 value\ucite{LEPEWWG9502} of 
$\Rb = 0.2219 \pm
0.0017$ of almost four standard deviations with respect to the SM
prediction, according to the 1996 result of $\Rb = 0.2179 \pm 0.0012$ 
presented in Warsaw, has reduced to less
than two standard deviations. 
In order to discuss the impact of $\Rb$ on the results for $\MH$, 
if $\Rb$ is included in the fits, we
recall that the SM prediction for $\Rb$ is (practically) independent of
the Higgs mass, but significantly dependent on $\Mt$. As the SM
prediction for $\Rb$ increases with decreasing mass of the top quark, $\Mt$, an
experimental enhancement of $\Rb$ effectively amounts\ucite{DSCW}
to imposing a low 
top-quark mass in fits of $\Mt$ and $\MH$, as soon as $\Rb$ is included in the
fits. Lowering the top-quark mass in turn implies a lowering of $\MH$ as a
consequence of the $(\Mt, \MH)$ correlation present in the theoretical values of
the other observables. Looking at fig. 5, we see that this effect of lowering
$\MH$ is not very significant with the 1996 value of $\Rb$ and the 1996
error in $\Mt$. The ``$\Rb$-crisis'' in the 1995 data, in contrast, led to a 
substantial decrease in the deduced value of $\MH$ to e.g. $\MH =
81^{+144}_{-52}$ GeV with $\swbar^2$ (LEP + SLD). As stressed in
\citere{DSCW},
this low value of $\MH$ had to be rejected, however, as the effective top-quark
mass induced by including $\Rb$ was substantially below the result from the
direct measurements at the Tevatron. Other consequences from the
``$\Rb$-crisis'', such as an exceedingly low value of $\alps \cong 0.100$ 
required upon allowing for a necessary
non-standard $Zb \bar b$ vertex, 
have also
gone away, and a very satisfactory and consistent overall picture of
agreement with Standard Model predictions has emerged. Speculations on the
existence of a ``leptophobic''\ucite{RE} or a ``hadrophilic'' extra
boson\ucite{RE,ALT,FR}, offered as
potential solutions
to the ``$\Rb$-crisis'', do not seem to be realized in nature. 

The delicate interplay of the experimental results for $\swbar^2 , \Rb$ and
$\Mt$ in constraining $\MH$ and the dependence of $\MH$ on $\alpha (\MZ^2)$ and
$\alps (\MZ^2)$
is visualized in the two-parameter $(\Mt , \MH)$ fits shown in
fig.~6. With its caption, fig.~6 is fairly self-explanatory. For a detailed
discussion we refer to the original papers\ucite{DSCW,DS}.
We only note the considerable dependence of the bounds resulting for $\MH$ on
whether the experimental value for $\Mt$ is included in the fit and the strong
dependence of $\MH$ on a $1\sigma$ variation of $\alpha (\MZ^2)$ and $\alps$.
{}Fig.~6 also shows that the SLD value of $\swbar^2$, when
taken by itself, would rule out an interpretation of the data in terms of the
standard Higgs mechanism, since the resulting Higgs mass, $\MH$, is much
below the lower bound of $\MH \geq 70$ GeV following from the direct 
Higgs-boson search at LEP. 

It is instructive to update our results (24) and (25) on $\MH$ from the
``leptonic sector'' (of $\swbar^2 , M_W , \Gamma_l$ together with
$m_t^{exp}$ and $\alpha(M^2_Z)$) on the basis of the `97 data, also shown
in tab. 1. One obtains
\beqar
\MH & = & 152^{+144}_{-~88}\GeV \quad {\rm using} \quad \swbar^2~({\rm
LEP + SLD})_{`97} = 0.23152\pm 0.00023, \nn \\
\MH & = & 265^{+208}_{-127}\GeV \quad {\rm using} \quad \swbar^2~({\rm LEP})_{`97} =
0.23196 \pm 0.00028 , 
\label{27}
\eeqar
thus implying $1 \sigma$ bounds of $\MH \lsim 300$ GeV and $\MH \lsim 470$
GeV, using $\swbar^2 ({\rm LEP + SLD})$ and $\swbar^2 ({\rm LEP})$,
respectively, and
\beqar
\MH & \lsim & 430 \GeV (95 \% C.L.) \quad {\rm using} \quad \swbar^2~({\rm
LEP + SLD})_{`97}, \nn \\
\MH & \lsim & 680 \GeV (95 \% C.L.) \quad {\rm using} \quad \swbar^2~({\rm
  LEP})_{`97}.
\label{28}
\eeqar
The somewhat lower values of $\MH$ extracted from the `97 data compared 
with the `96 data are largely due to an increase of the world average
value of $M_W$ by about 70 MeV (compare tab.~1). The 95 \% C.L. bound
of $M_H \lsim 430 \GeV$ (`97 data) from (28) is consistent with the
bound of $M_H \lsim\break 420 \GeV$ (`97 data) given in Ref. [4] in an
``all-data'' fit
which includes the hadronic sector.

\section{Production of \boldmath{W$^+$W$^-$} at LEP2}

In connection with the discussion of the coupling parameter $\Delta y$ in
sect.~3, we stressed that the agreement with the LEP1 data at the Z provides
convincing {\it indirect} experimental evidence for the non-Abelian couplings of the
Standard Model. More {\it direct, quantitative} information can be deduced from future
data on $e^+ e^- \rightarrow W^+ W^-$.

I start by quoting my dinstinguished  
late friend J.J. Sakurai. 
In his characteristic way of looking at physics, he said\ucite{SAK}:

\begin{quote}
``To quote Weinberg [Rev. Mod. Phys. {\bf 46} (1974) 255]
\begin{quote}
`Indeed, the best way to convince oneself that gauge theories may have something
to do with nature is to carry out some specific calulation and watch the
cancellations before one's very eyes'.
\end{quote} 
Does all this sound convincing? In any
case it would be fantastic to see how the predicted cancellations take place
{\it experimentally} at colliding beam facilities - LEPII? - in the 200 to 300
GeV range.'' 
\end{quote}
Unfortunately, J.J.\ was overly optimistic concerning the energy range of LEP2. 
My remark will be brief, and essentially consists of showing two figures.
The first figure will show our simulation on the accuracy to be expected
when extracting trilinear vector-boson couplings
from measurements of the reaction $e^+ e^- \rightarrow W^+ W^-$ at LEP2. 
The second figure will show the first experimental results obtained at LEP2. Restricting
ourselves to dimension-four, P- and C-conserving interactions, the general
phenomenological Lagrangian for trilinear vector boson couplings\ucite{BKRS}
\begin{eqnarray}
{\cal L}_{int}&=&-ie[A_\mu(W^{-\mu\nu}W^+_\nu-W^{+\mu\nu}W^-_\nu)
+F_{\mu\nu}W^{+\mu}W^{-\nu}]
\nonumber\\ &&
-iex_\gamma F_{\mu\nu}W^{+\mu}W^{-\nu}
\label{23}
\\
&&-ie\biggl(\frac{c_W}{s_W}+\delta_Z\biggr)
[Z_\mu(W^{-\mu\nu}W^+_\nu-W^{+\mu\nu}W^-_\nu)
+Z_{\mu\nu}W^{+\mu}W^{-\nu}]
\nonumber\\ &&
-iex_Z Z_{\mu\nu}W^{+\mu}W^{-\nu}
\nonumber
\end{eqnarray}
is obtained by supplementing the trilinear interactions of the SM with an
additional anomalous magnetic-moment coupling of strength $x_\gamma$, by
allowing for arbitrary normalization of the Z~coupling via $\delta_Z$, and
by adding an additional anomalous weak magnetic dipole coupling of the Z of
strength $x_Z$.
Compare \citere{gks2} for a representation of the effective Lagrangian \refeq{23}
in an SU(2)$\times$U(1) gauge-invariant form.  
The SM corresponds to $x_\gamma = \delta_Z = x_Z = 0$.

\begin{figure}
\begin{center}
\begin{tabular}{c@{~}c}
\epsfig{file=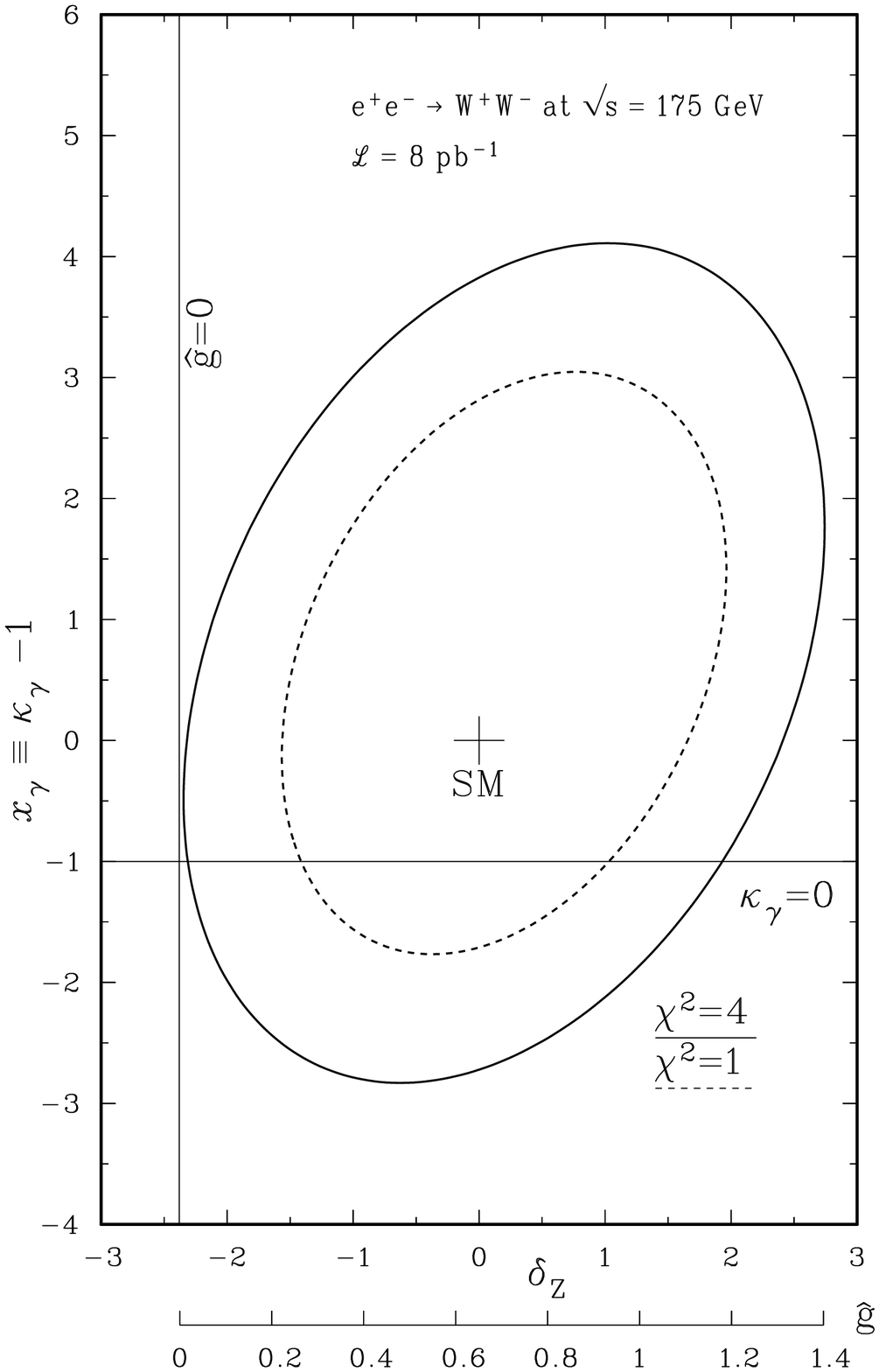,width=6.2cm,
bbllx=40,bblly=40,bburx=490,bbury=710}&
\epsfig{file=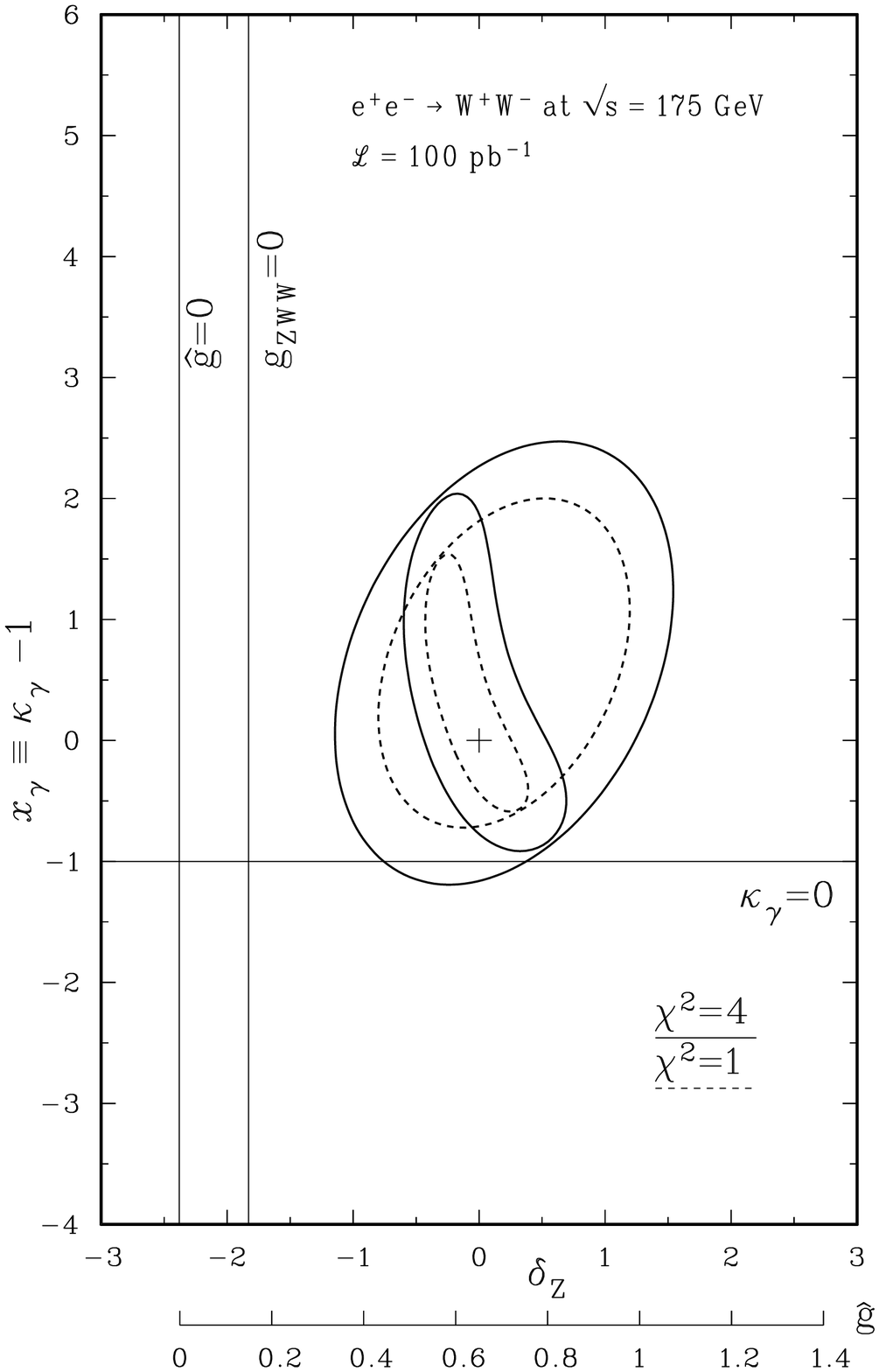,width=6.2cm,
bbllx=40,bblly=40,bburx=490,bbury=710}\cr
a&b
\end{tabular}
\end{center}
\caption{\it a: Detecting the existence of a non-Abelian 
vector-boson coupling, 
\protect{$\hat{g}\neq 0$}, at LEP~2.
b: Detecting a non-zero anomalous magnetic dipole moment,
\protect{$\kappa_\gamma\neq 0$}, of the \protect{$W^{\pm}$} at LEP~2.}
\label{anom}
\end{figure}

Non-vanishing values of $x_\gamma$ parametrize deviations of the magnetic
dipole moment, $\kappa_\gamma$, from its SM value of $\kappa_\gamma = 1$, as
according to \refeq{23}, 
\beq 
x_\gamma \equiv \kappa_\gamma - 1.
\label{24}
\eeq
We note that $\kappa_\gamma = 1$ corresponds to a gyromagnetic ratio , $g$,
of the $W$ of magnitude $g = 2$ in units of the particle's Bohr-magneton
$e/2 \MW$, while $\kappa_\gamma = 0$ corresponds to $g = 1$ as obtained for a
classical rotating charge distribution. The weak dipole coupling, $x_Z$, may be
related to $x_\gamma$ by imposing ``custodial'' SU(2) symmetry via\ucite{MSS}
\beq
x_Z = - \frac{s_W}{c_W} x_\gamma, 
\label{25}
\eeq
thus reducing the number of free parameters to two independent ones in \refeq{23}.
Relation \refeq{25} follows from requiring the absence of an SU(2)-violating
interaction term solely among the members of the SU(2) triplet, $W^3_{\mu\nu}
W^{+\mu}\break W^{-\nu}$, when rewriting the 
Lagrangian in the $BW^3$ base (or the 
$\gamma W^3$ base). This requirement is motivated by the validity of SU(2)
symmetry for the vector-boson mass term, i.e. 
from the observation that the deviation of the experimental
value for $\Delta x$ from $\Delta x = 0$ in sect.~1.3 is fully explainable by
radiative corrections, thus ruling out a violation of ``custodial'' SU(2)
symmetry by the vector boson masses at a high level of accuracy. 

We also note the relation of $\delta_Z$ to the weak gauge coupling $\hat g$
describing the trilinear coupling between $W^0$ and $W^\pm$ in the $BW^0$ (or
$\gamma W^3$) base, 
\beq
e \delta_Z \equiv g_{ZWW} - e \frac{c_W}{s_W} = \frac{\hat g}{c_W} -
\frac{e}{s_W c_W} .
\label{30}
\eeq
The SM corresponds to $\hat g = e / s_W$. 

\begin{figure}
\begin{center}
\epsfig{file=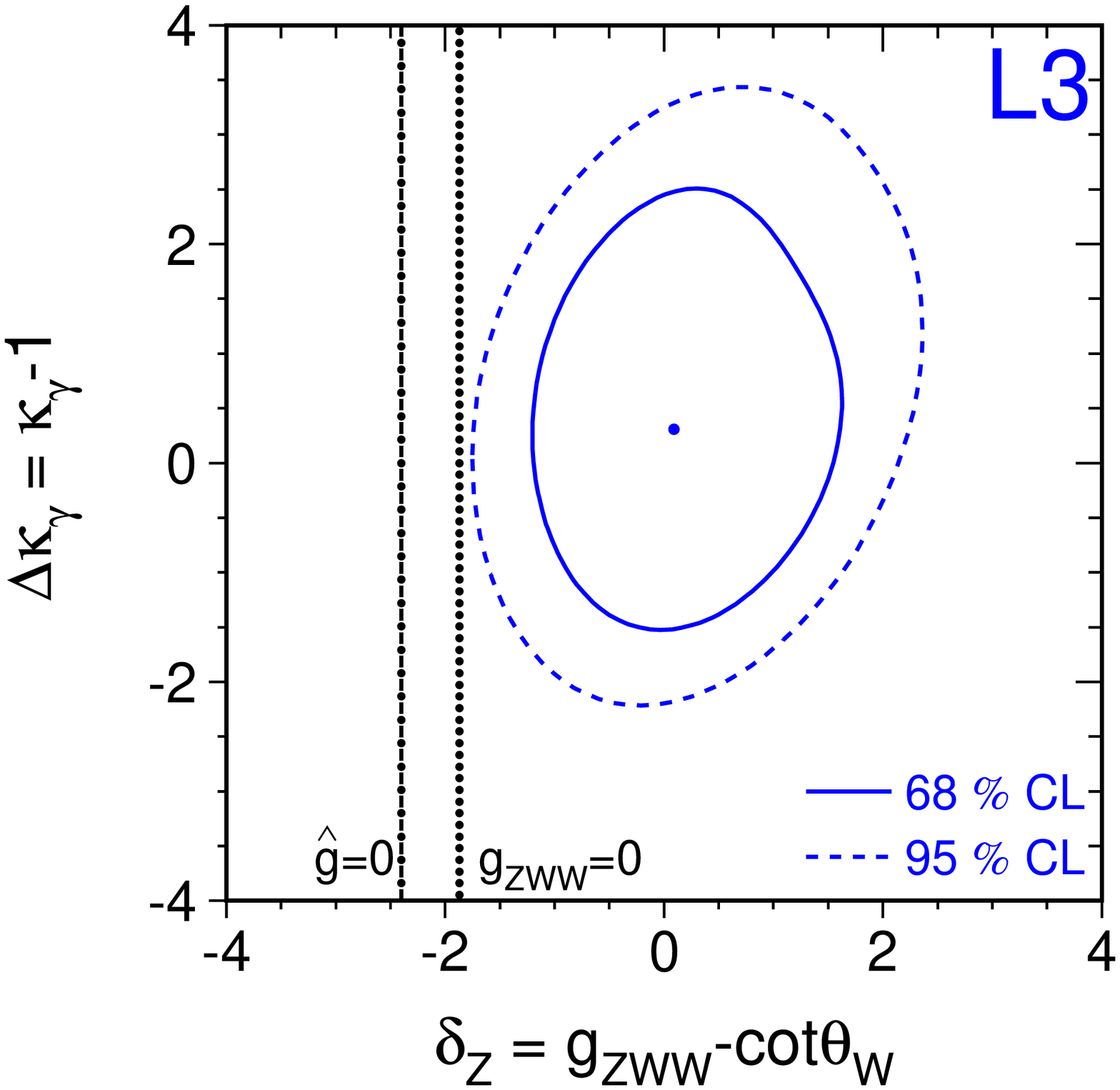,width=12cm,}
\end{center}
\caption{\it Bounds on $\Delta \kappa_\gamma \equiv X_\gamma$ and $\delta_Z$
obtained [43] by the L3 Collaboration at LEP.}
\efi
{}Figs. \ref{anom}a and \ref{anom}b from 
\citere{KS} are based on the assumption 
that future data on $e^+ e^-
\rightarrow W^+ W^-$ at an energy of 175 GeV will agree with SM predictions
within errors. Under this assumption, fig.~\ref{anom}a shows that an integrated luminosity
of $8 pb^{-1}$, corresponding to a few weeks of running at $175\GeV$
will be sufficient to
provide {\it direct} experimental evidence for the existence of a non-vanishing
coupling of the non-Abelian type, $\hat g \not= 0$,  
among the members of the vector-boson triplet
(at 95\% C.L.). Likewise, according to fig.~\ref{anom}b, 
an integrated luminosity of $100 pb^{-1}$,
corresponding to about seven months of running at LEP2, will provide direct
experimental evidence for a non-vanishing anomalous 
magnetic moment of the W boson
(at 95\% C.L.), $\kappa_\gamma \not= 0$.

Figure 8 finally shows the experimental result\ucite{L3} recently obtained by
the L3 collaboration. The data at 95 \% C.L. indeed rule out a vanishing
weak (trilinear) coupling, $\hat g$, among the members of the $W^0, W^\pm$
triplet as well as a vanishing of the $ZW^+W^-$ coupling, $g_{ZW^+W^-}$.

\section{Conclusions}

Let me conclude as follows:
\begin{itemize}
\item[i)]
The Z~data and the W-mass measurements require electroweak corrections
beyond fermion-loop contributions to the vector-boson propagators. 
\item[ii)]
In the Standard Model such corrections are provided by bosonic loops. The
dominant bosonic correction needed for agreement with the data can be traced
back to the difference in scale between $\mu$ decay, entering via $G_\mu$, and
W or Z decay. While not being sensitive to the Higgs mechanism, these
bosonic corrections depend on the non-Abelian couplings among the vector
bosons. The data accordingly ``see'' the non-Abelian structure of the Standard
Model.  
\item[iii)]
The bounds on the mass, $\MH$, of the Higgs scalar are most reliably derived
from the reduced set of data containing $\swbar^2 , \MW , \Gl , \Mt^{{\rm
exp}}$ and $\alpha (\MZ^2)$ besides $\MZ$ and $G_\mu$. At 95\% C.L. the 1996 set
of data implies $\MH \lsim 550$ GeV and $\MH \lsim 800$ GeV, 
depending on whether
$\swbar^2$(LEP+SLD) or $\swbar^2$(LEP) is used as input. The `97 data
improve these bounds to $M_H \lsim 430 \GeV$ and $M_H \lsim 680 \GeV$,
respectively.
These bounds are quite remarkable, as for the first time they seem to fairly
reliably predict a Higgs mass in the perturbative region of the SM.
\item[iv)]
Since the ``$\Rb$-crisis'' has meanwhile been resolved by our experimental
collegues, there is now perfect overall
agreement with the predictions of the SM, even upon including hadronic $Z$
decays in the analysis. The strong coupling, $\alps (\MZ^2)$, obtainable from
the hadronic Z-decay modes, comes out consistently with the event-shape
analysis. Various speculations on ``hadrophilic'' or ``leptophobic'' bosons do
not seem to be realized in nature.    
\item[v)]
{}The experiments at LEP2 on $e^+ e^- \rightarrow W^+ W^-$ show
{\it direct} experimental evidence for the existence of non-vanishing
couplings of non-Abelian type among the vector bosons. 
\item[vi)]
The available data by themselves do not discriminate a MVB from the
Standard Theory based on the Higgs mechanism. The issue of mass generation will remain
open until the Higgs scalar will be found - or something else?
\end{itemize}

\bigskip 
\noindent{\large\bf Acknowledgement}\\[1mm]
It is a pleasure to thank Misha Bilenky, Stefan Dittmaier, Carsten 
Grosse-Knetter, 
Karol Kolodziej, Masaaki Kuroda, Ingolf Kuss and Georg Weiglein for a fruitful 
collaboration on various aspects on the theory of  
electroweak interactions.

\end{document}